\documentclass[pra,twocolumn,
                                groupedaddress,
                                showpacs,
                                ]{revtex4}
\usepackage{graphicx}
\usepackage{amsmath}
\usepackage{amsfonts}
\usepackage{amssymb}
\usepackage{hyperref}
\usepackage{epstopdf}

\newcommand{\beq}{\begin{equation}}
\newcommand{\eeq}{\end{equation}}

\usepackage{placeins} 
\usepackage{bm}         

\begin{document}

\title{Thermodynamics and dynamics of atomic self-organization in an optical cavity}

\author{Stefan Sch\"utz} 
\affiliation{Theoretische Physik, Universit\"at des Saarlandes, D-66123 Saarbr\"ucken, Germany} 

\author{Simon B. J\"ager} 
\affiliation{Theoretische Physik, Universit\"at des Saarlandes, D-66123 Saarbr\"ucken, Germany}

\author{Giovanna Morigi} 
\affiliation{Theoretische Physik, Universit\"at des Saarlandes, D-66123 Saarbr\"ucken, Germany} 

\date{\today}

\begin{abstract}
Pattern formation of atoms in high-finesse optical resonators results from the mechanical forces of light associated with superradiant scattering into the cavity mode. It occurs when the laser intensity exceeds a threshold value, such that the pumping processes counteract the losses. We consider atoms driven by a laser and coupling with a mode of a standing-wave cavity and describe their dynamics with a Fokker-Planck equation, in which the atomic motion is semiclassical but the cavity field is a full quantum variable. The asymptotic state of the atoms is a thermal state, whose temperature is solely controlled by the detuning between the laser and the cavity frequency and by the cavity loss rate. From this result we derive the free energy and show that in the thermodynamic limit selforganization is a second-order phase transition. The order parameter is the field inside the resonator, to which one can associate a magnetization in analogy to ferromagnetism, the control field is the laser intensity, however the steady state is intrinsically out-of-equilibrium. In the symmetry-broken phase quantum noise induces jumps of the spatial density between two ordered patterns: We characterize the statistical properties of this temporal behaviour at steady state and show that the thermodynamic properties of the system can be extracted by detecting the light at the cavity output. The results of our analysis are in full agreement with previous studies, extend them by deriving a self-consistent theory which is valid also when the cavity field is in the shot-noise limit, and elucidate the nature of the selforganization transition. 
\end{abstract}


\maketitle

\section{Introduction}

There is ample experimental evidence that electromagnetic fields can cool matter to ultralow temperatures \cite{Nobel,LaserCooling,Aspelmeyer}. This is achieved by tailoring scattering processes, so that the frequency of the emitted photon is on average larger than that of the absorbed one, the energy balance being warranted by the mechanical energy which is exchanged between matter and light  \cite{Wineland:1978,Stenholm:1986}. When atoms or molecules interact with high-finesse optical resonators, these processes can be tailored using the strong coupling with the cavity field \cite{Horak:1997,Pinkse:2000,Kimble:2000,Vuletic:2000,Pinkse:2004,Domokos:2003,Ritsch,Kampschulte:2014}.  

\begin{widetext}
\begin{figure*}[hbtp]
\begin{center}
\includegraphics[width=0.9\textwidth]{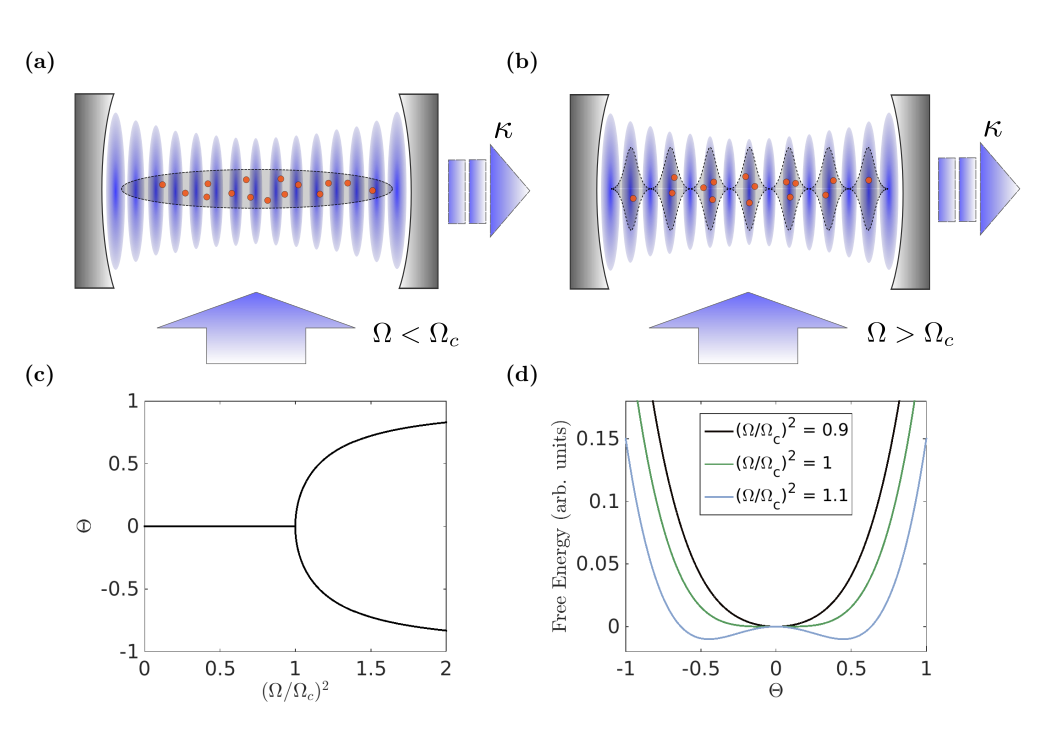}
\caption{(color online) (a) Atoms in a standing-wave cavity and driven by a transverse laser can spontaneously form ordered patterns (b) when the laser intensity $\Omega$ exceeds a threshold value $\Omega_c$, which depends on the rate of photon losses, here due to cavity decay at rate $\kappa$. In this regime the atoms experience a long-range interaction mediated by the cavity photons and their motion becomes strongly correlated. (c) Spatial ordering of atoms is described by the parameter $\Theta$, which characterizes the localization of the atoms within the standing-wave mode of the cavity and is proportional to the cavity field. This parameter undergoes a bifurcation at $\Omega=\Omega_c$, corresponding to two different stable patterns. The values it takes are the minima of an effective Landau potential, displayed in (d) for some values of $\Omega$, demonstrating that selforganization is a second-order phase transition. See text for details.}
\label{Fig:1}
\end{center}
\end{figure*}
\end{widetext}

A peculiar aspect of light-matter interaction inside optical cavities are the long-range interactions between the atoms, which are mediated by multiple scattering of photons \cite{ODell,Rempe}. The onset of this behaviour is observed when the system is driven by external pumps, whose strength overcomes the loss rate. Some prominent examples are optomechanical bistability \cite{Bistability,Ritter:2009}, synchronization \cite{CARL_Zimmermann}, and spontaneous spatial ordering \cite{Domokos:2002,Black:2003,Baumann:2010,Chang,Oppo,Ritsch}. Among several setups, spontaneous pattern formation in standing-wave and single-mode cavities has been object of several theoretical and experimental studies \cite{Ritsch}. This phenomenon occurs when the  atoms are confined within the resonator and are transversally driven by a laser, and consists in the formation of atomic gratings that maximize coherent scattering of laser photons into the cavity mode, as sketched in Fig. \ref{Fig:1}(a) and (b). These "Bragg gratings" are stably trapped by the  mechanical effects of the light they scatter, provided that the laser compensates the cavity losses so that the number of intracavity photons is sufficiently large. It takes place when the laser intensity, pumping the atoms, exceeds a threshold value depending, amongst others, on the rate of photon losses and on the number of atoms \cite{Domokos:2002,Ritsch}. This behaviour was first  predicted in Ref. \cite{Domokos:2002}, and experimentally demonstrated in several settings, which majorly differ from the initial temperature of the atomic ensemble: In Refs. \cite{Black:2003,Arnold:2012} the atoms were cooled by the mechanical effects of the photons scattered into the resonator, while in Refs. \cite{Baumann:2010,Mottl:2012} the atoms initially formed a Bose-Einstein condensate, and the mechanical effects of light were giving rise to conservative forces. As a consequence, matter-wave coherence was preserved during the experiment. In this regime, the transition to selforganization can be cast in terms of the Dicke phase transition \cite{Domokos:Dicke}.

In this work we theoretically analyse the dynamics leading to the formation of spatial structures and their properties at the asymptotics. Our analysis is based on a semiclassical treatment, and specifically on a Fokker-Planck equation (FPE) derived when the atoms are classically polarizable particles and  their center-of-mass motion is along one dimension \cite{Schuetz:2013}. The cavity field, instead, is a full quantum variable, which makes our treatment valid also in the shot-noise limit \cite{Schuetz:2013} and describes parameter regimes that are complementary to those of the model in Ref. \cite{Horak:2001}, where the field is a semiclassical variable. Our formalism permits us, in particular, to consistently eliminate the cavity variables from the equations of motion of the atoms, and to analyse the properties of the cavity field across the selforganization threshold, where the intracavity field is characterized by large fluctuations. 

This work extends and complements the study presented in Ref. \cite{Schuetz:2014}. In particular, we perform a detailed analysis of the stationary state and obtain an analytic expression, which allows us to determine the phase diagram of the transition as a function of the relevant parameters. Drawing from this result, in addition, we show that the onset of selforganization in spatially ordered patterns is a second-order phase transition, associated with a symmetry breaking in the phase of the intracavity field. This allows us to verify conjectures on the nature of the selforganization transition, previously discussed in Refs. \cite{Asboth:2005,DallaTorre:2011,Niedenzu:2012}. 
We further analyse in detail the effects of the nature of the long-range interactions mediated by the photons and  report on several features which are analogously found in the Hamiltonian-Mean-Field (HMF) model, the workhorse of the statistical physics with long-range interactions \cite{Campa:2009}. This article is the first of a series of works devoted to the semiclassical theory of selforganization. 

In the present work we analyse the thermodynamics of selforganization and the dynamics at the asymptotics, while in following articles we  investigate the dynamics following sudden quenches across the phase transition \cite{Schuetz:2015:2} and compare our analysis with a mean-field model, that discards some relevant effects of the long-range correlations \cite{Jaeger:2015}. This manuscript is organized as follows. In Sec. \ref{Sec:Model} the Fokker-Planck Hamiltonian at the basis of our analysis is reported and discussed. In Sec. \ref{Sec:Steady} the stationary properties of the distribution function are characterised both analytically as well as numerically. In Sec. \ref{Subsec:cav} the correlation functions of the light at the cavity output are determined. The conclusions are drawn in Sec. \ref{Sec:Conclusions}, while the Appendices report details of analytical calculations and of the numerical program, that is used to simulate the FPE. 

\section{Model}
\label{Sec:Model}

The dynamics of $N$ atoms or molecules of mass $m$ inside a single-mode standing-wave cavity is analysed when the particles are transversally illuminated by a laser field, as illustrated in Fig. \ref{Fig:1}(a). Laser and cavity couple to a dipole transition of the scatterers and are assumed to be sufficiently far-off resonance so that the coupling with the internal degrees of freedom is described by the particles polarizability. From now on we will assume that the particles are atoms, but the treatment in this paper can be extended to any ensemble of linearly polarizable particle that can be confined within the optical resonator \cite{Polarizable}.

In this regime the atoms scatter all coherently and the cavity field $E_c$ is the sum of the fields each atom scatters. We assume that the atoms center-of-mass motion is  confined along the cavity axis, which coincides with the $x$-axis (we disregard their motion in the transverse plane), and that the atoms are uniformly illuminated by the laser field. Denoting the atomic position by $x_j$ and the cavity-mode function by $\cos(kx)$,  with $k$ the wave number, then $E_c\propto N\Theta$, where
\beq
\label{Theta}
\Theta=\frac{1}{N}\sum_j\cos(kx_j)\,
\eeq
measures the ordering of the atoms within the cavity standing-wave. For $N\gg 1$, when the atoms are uniformly distributed, $\Theta\sim 0$ and the field within the cavity vanishes. The intracavity intensity is maximal when the positions are such that $\cos(kx_j)=1$ (even pattern) or $\cos(kx_j)=-1$ (odd pattern), namely, when the atoms form Bragg gratings, see Fig. \ref{Fig:1}(b). These gratings are the two possible stable configurations the atoms can form when the laser pump is above threshold, as shown in Fig. \ref{Fig:1}(c).

The formation and stability of the Bragg gratings is determined by the mechanical effects of photon scattering on the atoms. In this section we report the basic equations describing the dynamics of the coupled systems, as well as the assumptions that lead to a Fokker-Planck equation (FPE) governing the semiclassical trajectories of $N$ atoms inside the single-mode resonator \cite{Schuetz:2013}. The FPE is derived under the assumption that the atomic motion is at all times in the semiclassical regime, while the cavity field adjusts quasi-instantaneously to the atomic density distribution. In this limit, using a perturbative treatment the cavity field can be eliminated by the equations of motion of the atoms external degrees of freedom \cite{Dalibard:1985}. The readers interested in the detailed derivation of the FPE from the full quantum master equation of atoms and cavity are referred to Refs. \cite{Dalibard:1985,Schuetz:2013}. An alternative FPE, where fluctuations of the intracavity field are treated semiclassically but no time-scale separation between atoms and cavity dynamics is assumed, is derived in Ref. \cite{Horak:2001}.

\subsection{The cavity field}

In our treatment the cavity field is a quantum variable. We report its equation of motion in the limit in which the atoms constitute a non-saturated medium and their internal atomic transitions are described by the polarizability.  Our starting point is the Heisenberg-Langevin equation for operator $\hat{a} (t)$, which annihilates a cavity photon at frequency $\omega_c$ and wave number $k$. The equation is reported in the reference frame rotating at the laser frequency $\omega_L$ and reads \cite{Fernandez-Vidal:2010}
\begin{eqnarray}
\label{a:field}
\frac{\partial}{\partial t} \hat{a} (t) 
&=&- \left[ \kappa- i (\Delta_c - NU \hat{\mathcal B}(t))\right] \hat{a} (t) \\
& &- i N S \hat\Theta (t) +\hat\xi (t)\,,\nonumber
\end{eqnarray}
where $\Delta_c=\omega_L-\omega_c$ is the detuning of the laser from the cavity frequency, $\hat\xi(t)$ is the Langevin force with $\langle \hat\xi(t')\hat\xi^\dagger(t)\rangle=2\kappa \delta(t-t')$ and $\kappa$ the cavity decay rate. The cavity field is a function of the two operators $\hat{\mathcal B}(t)$ and $\hat\Theta(t)$, which in turn are functions of the atomic positions $\hat x_j$ at time $t$. In detail, $U$ is a frequency, $U=g^2/\Delta_a$, where $g$ is the vacuum Rabi frequency at the antinodes of the cavity mode, $\Delta_a=\omega_L-\omega_a$ is the detuning of the laser frequency from the atomic transition resonance $\omega_a$, and operator $\hat{\mathcal B}$ is defined as
\begin{align}
\hat{\mathcal B}=\frac{1}{N}\sum_j\cos^2(k\hat x_j) \, ,
\label{eq:bunch_hat}
\end{align}
and takes on values between 0 and 1. Its expectation value $B=\langle \hat{\mathcal B}\rangle$ is the so-called bunching parameter \cite{Ritsch}. Operator $\hat\Theta(t)$ is the quantum variable corresponding to the order parameter in Eq. \eqref{Theta}. In Eq. \eqref{a:field} it is scaled by the frequency $S=\Omega g/\Delta_a$, which is proportional to the laser Rabi frequency $\Omega$ and corresponds to the scattering amplitude of a laser photon into the cavity mode by an atom at an antinode, with $S/U=\Omega/g$. Equation \eqref{a:field} shows that the pump on the cavity is maximum when $\langle \hat\Theta\rangle=\pm 1$, corresponding to the situation in which the atoms form Bragg gratings. Selforganization occurs when these gratings are mechanically stable, namely, when the mechanical effects of the scattered light stabilize the atoms in ordered structures, which in turn generate the field. In order to determine these dynamics one would need to solve the coupled equations of cavity and atomic motion. 

We can further simplify the problem by considering the regime in which the time scale over which the atomic motion evolves is much larger than the time scale determining the evolution of the cavity field.  This is typically fulfilled when $k \bar p/m\ll |\kappa+{\rm i}\Delta_c|$, where $\bar p=\sqrt{\langle \hat p^2 \rangle}$ is the variance of the atomic momentum (the mean value vanishes), under the condition that the coupling between cavity and atomic motion is sufficiently weak. This latter condition requires that \cite{Footnote}
\beq
\label{Eq:condition}
\sqrt{\omega_r}\sqrt{N}|S| \ll  | \Delta_c + i \kappa|^{3/2}\,,
\eeq 
where $\omega_r=\hbar k^2/(2m)$ is the recoil frequency, scaling the exchange of mechanical energy between photons and atoms. At zero order in this expansion  the cavity field operator depends on the instantaneous density and reads
\beq
\hat{a}_{\rm ad}(t) = \frac{N S \hat\Theta (t) }{\hat\Delta_c'(t) +i \kappa}\,, 
\label{a_ad}
\eeq
where the subscript indicates the adiabatic limit and we omitted to report the noise term. Operator $\hat\Delta_c'$ is defined as
\beq
\label{Delta:c}
\hat\Delta_c^\prime=\Delta_c-UN\hat{\mathcal B}\,.
\eeq 
Its mean value vanishes for certain density distributions, giving rise to resonances. For  $|NU|> \kappa$ small changes of $\Delta_c$ about the resonance can induce large variations of the field, resulting in the appearance of optomechanical bistable behaviour \cite{Bistability,Ritter:2009,Larson:2008}. In this paper we focus on the regime in which $|NU| \ll \kappa$, and treat this as a small parameter on the same footing as the retardation term. In this limit, the field, including the diabatic corrections, reads 
\begin{eqnarray}
\hat{a}(t) = \frac{N S \hat{\Theta} (t)}{ \Delta_c + i \kappa} \big[ 1 + \frac{NU}{\Delta_c + i \kappa} \hat{\mathcal{B}}(t)\big] + \hat{a}_{\rm ret}(t)\,,
\label{a(t)}
\end{eqnarray}
where
\beq
\hat{a}_{\rm ret}(t)= \frac{ i N S}{(i \Delta_c - \kappa)^2}\dot{\hat\Theta}
\label{a_retard}
\eeq
accounts for retardation effects and depends on the time derivative of operator $\hat\Theta$, Eq. \eqref{Theta}. The derivative in particular takes the form
$$\dot{\hat\Theta} =- \frac{1}{2N} \sum_j \big( \sin(k\hat{x}_j(t)) \frac{k\hat{p}_j(t)}{m} + \frac{k\hat{p}_j(t)}{m}\sin(k\hat{x}_j(t)) \big)\,, $$ 
and shows that the diabatic correction scales with $(k \bar p /m)/| \kappa + {\rm i} \Delta_c|$. When this parameter is small, then, one can perform a coarse-graining for the atomic motion, over which the cavity field fast relaxes. 

It is also useful to discuss the mean number of photons inside the resonator. In the adiabatic limit it is given by 
\beq
\langle \hat n\rangle_{t,{\rm ad}} =N\bar n\langle \hat \Theta^2\rangle_t\,,
\label{photons_adiabatic}
\eeq
which is valid in zero order in the delay time. For later convenience, we introduced the dimensionless quantity
\beq 
\label{bar:n}
\bar n=\frac{NS^2}{\Delta_c^{ 2}+\kappa^2}\,,
\eeq
such that $N \bar n$ gives the maximum intracavity photon number, corresponding to the value $\langle \Theta^2\rangle_t=1$, namely, when the atoms form a perfectly-ordered Bragg grating. The average photon number  can be different from zero also when the field inside the resonator has vanishing mean expectation value, since in this case it is proportional to the fluctuations of the order parameter.

\subsection{Fokker-Planck equation for $N$ atoms}

An equation for the motion of the $N$ atoms within the resonator is derived under the assumption that at all times the atomic momentum distribution has width $\Delta p=\bar p$ which is much larger than the quantum of linear momentum $\hbar k$ the atom exchanges with the individual photons (but sufficiently small so that the atoms are within the velocity capture range \cite{Domokos:2003}). This assumption is valid for cavities whose decay rate $\kappa$ exceeds the recoil frequency $\omega_r$: $\omega_r\ll \kappa$. In fact, we will show that $\kappa$ determines the minimum stationary width of the momentum distribution. This regime is encountered in several existing experiments \cite{Black:2003,Ritter:2009,Arnold:2012}. We note that, with this assumption, the requirement of time-scale separation between cavity and motion is fulfilled, since the inequality $k \bar p/m\ll \kappa$ is consistent with $\omega_r\ll \kappa$ after using $\bar p^2 /2m=\hbar \kappa/2$.

Reference \cite{Schuetz:2013} reports the detailed steps that lead to the derivation of a FPE for the distribution $f(\bm{x},\bm{p},t)$ of the $N$ atoms positions and momenta $\bm{x}~=~(x_1, x_2, ..., x_N)$ and $\bm{p}~=~(p_1, p_2, ..., p_N)$. The FPE can be cast in the form
\beq
\label{FPE:0}
\frac{\partial f}{\partial t} = - \sum_i \frac{p_i}{m} \frac{\partial}{\partial x_i} f+S^2 L\,f\,,
\eeq
where $f\equiv f(\bm{x},\bm{p},t)$. The Right-Hand Side (RHS) separates the ballistic motion from the term proportional to the scattering rate $S$ and describes the dynamics due to the mechanical effects of light. This latter term specifically reads
\begin{eqnarray}
L\,f&=&- \sum_i \frac{\partial}{\partial p_i} F_0(\bm{x})\sin(kx_i)~ f \label{L:f}\\  
&&- \sum_{i,j}  \frac{\partial}{\partial p_i} \Gamma_0(\bm{x})\sin(kx_i) \sin(kx_j) p_j~f  \nonumber\\
&&+  \sum_{i,j} \frac{\partial^2}{\partial p_i \partial x_j}\eta_{0}(\bm{x})\sin(kx_i) \sin(kx_j)~f\nonumber\\
&&+\sum_{i,j}  \frac{\partial^2}{\partial p_i \partial p_j}D_0(\bm{x})\sin(kx_i) \sin(kx_j)~f \nonumber \\  
&&+\frac{\gamma'}{2}\sum_{i}   \frac{\partial^2}{\partial p_i^2}\mathcal{D}^{\text{sp}}(x_i)~f \nonumber \,.
\end{eqnarray}
Here, the first term on the RHS describes the dispersive force associated with scattering of laser photons into the resonator, where
\beq
\label{F:0}
F_0(\bm{x})  = (\hbar k)  \frac{2 \Delta_c'}{\Delta_c'^2 + \kappa^2}(1+ \delta_F ) N\Theta \,.
\eeq
Its amplitude is proportional to the order parameter $\Theta$, Eq. \eqref{Theta},  which is the Wigner representation of operator $\hat \Theta$  \cite{Schuetz:2013}. Its sign is also determined by the frequency shift of the cavity frequency $\Delta_c^\prime({\bm x})$ from the laser, which takes the same form as in Eq. \eqref{Delta:c}, now with the corresponding Wigner form for operator $\hat{\mathcal B}$. Coefficient $\delta_F$ is a small correction for the parameter regime we consider, its general form is given in Appendix \ref{App:Parameter}. The same applies for the coefficients $\delta_j$ ($j=\Gamma,\eta,D$) appearing in the other terms we specify below.

The second term on the RHS of Eq. \eqref{L:f} describes the damping force due to retardation between the scattered field and the atomic motion. It depends on the atomic momentum and is scaled by the function
\beq
\label{Gamma:0}
\Gamma_0(\bm{x}) = \omega_r  \frac{8 \Delta_c' \kappa}{(\Delta_c'^2 + \kappa^2)^2}(1+\delta_\Gamma)\,.
\eeq
The third summand is due to the anharmonicity of the cavity optical lattice. The function scaling this term has the form
\beq
\label{eta:0}
\eta_0(\bm{x}) = 2 \hbar \omega_r \frac{ (-\Delta_c'^2 + \kappa^2)}{(\Delta_c'^2 + \kappa^2)^2}(1+\delta_\eta)
\eeq
and vanishes when $\Delta_c'=\pm\kappa$. 

The last two terms describe diffusion. In particular, the one scaled by the function
\beq
\label{D:0}
D_0(\bm{x}) = (\hbar k)^2 \frac{ \kappa}{\Delta_c'^2 + \kappa^2} (1+\delta_D)
\eeq
corresponds to the diffusion associated with global fluctuations of the cavity field and is characterized by long-range correlations, while
the term with coefficient $\mathcal{D}^{\text{sp}}(x_i) $  is instead due to spontaneous emission of a photon outside the resonator with $\gamma'=\gamma g^2 / \Delta_a^2$, where $\gamma$ is the decay rate of the excited state. It is the sole term which acts locally, and the dynamics it implies does not establish correlations between the atoms. Its explicit form is reported in Appendix \ref{App:Parameter}. 

\subsection{Dynamics away from the bistable regime}

Equation \eqref{FPE:0} describes the coherent and dissipative dynamics associated with the mechanical effects of light on the atomic motion. In this work we will assume that $\gamma'$ is much smaller than the other rates and discard the effect of spontaneous decay in the dynamics, so that losses are due to cavity decay. As far as it concerns the terms due to the cavity, we note their nonlinear dependence on the bunching parameter, which appears in the denominator of all coefficients and gives rise to bistable behaviour.  Here, we focus on the regime in which $|NU| \ll \kappa$. In this regime the dispersive forces due to the mechanical effects of light in leading order are due to scattering of laser photons into the cavity. In this limit, we choose detunings $|\Delta_c|\sim \kappa$ so that the motion is efficiently cooled, as we show below. Correspondingly, the coefficients of the functional in Eq. \eqref{L:f} are modified so that $\Delta_c'\simeq \Delta_c$ and the functions $\delta_F,\delta_\eta,\delta_\Gamma,\delta_D\approx 0$. More precisely, we perform an expansion in first order in $N|U|/\kappa$. In this limit, the Fokker-Planck equation, Eq. \eqref{FPE:0}, can be cast in the form 

\begin{widetext}
\begin{align}
\label{FPE}
&\partial_t f+\{f,H\}+\bar n\frac{NU}{\Delta_c}L_1f
=- \bar n\Gamma\sum_{i} \sin(kx_i) \partial_{p_i}\frac{1}{N} \sum_j \sin(kx_j) \left( p_j + \frac{m}{ \beta} \partial_{p_j}+\frac{\bar\eta}{\beta}\partial_{x_j}\right)f\,,
\end{align}
\end{widetext}
where all terms due to the coupling with the light scale with $\bar n$, given in Eq. \eqref{bar:n}. In detail, the Left-Hand Side (LHS) collects the hamiltonian terms, expressed in terms of Poisson brackets with Hamiltonian 
\begin{eqnarray}
\label{Hmf_self}
H=\sum_j \frac{p_j^2}{2m}+  \hbar\Delta_c\bar n N\Theta^2\,,
\end{eqnarray}
as well as the terms scaling with $U$, summarized in the functional $L_1$, whose detailed form is given in App. \ref{App:Parameter}. The RHS reports terms of different origin, which can be classified as damping, diffusion, and a third term which scales cross-derivatives in position and momentum. In the order of this list, they are scaled by the coefficients
\begin{eqnarray}
&&\Gamma=8\omega_r\kappa\Delta_c/(\Delta_c^2 + \kappa^2)\,,\\
&&\beta=-4\Delta_c/\hbar/(\Delta_c^2 + \kappa^2)\,, \label{eq:hbar_beta}\\
&&\bar\eta=\frac{\kappa^2-\Delta_c^2}{\kappa(\Delta_c^2 + \kappa^2)}\,.
\end{eqnarray}
We remark that the term in the FPE scaled by parameter $\bar\eta$ was already found in the derivation of Ref. \cite{Dalibard:1985}. While its effect is to date not well understood, we checked that for the parameters we consider it gives rise to small corrections in the quantities we evaluate. In the mean-field treatment it can be cast in terms of a correction of the effective mean-field potential the atoms experience. In that limit it induces a shift to the critical value of the pump strength at the selforganization transition \cite{Jaeger:2015}.

\subsection{Long-range correlations}

Let us now make some preliminary remarks on the FPE discussed this far. We first focus on the Hamiltonian term, Eq. \eqref{Hmf_self}. In addition to the kinetic energy this contains  
the cavity-mediated potential, which has been obtained in zero order in the retardation time. Its sign is determined by the sign of the detuning $\Delta_c$: When $\Delta_c<0$,  the formation of Bragg gratings, which maximizes the value of $|\Theta|$, is energetically favoured. Thus, Eq. \eqref{Hmf_self} summarizes in a compact way a property which was observed in several previous works  \cite{Domokos:2002,Black:2003,Asboth:2005,Schuetz:2014}. 

We note that the Hamiltonian in Eq. \eqref{Hmf_self} exhibits several analogies with the Hamiltonian Mean Field (HMF) model \cite{Campa:2009}, whose Hamiltonian reads
\beq
H_{\rm MF}=\sum_j \frac{p_j^2}{2m}+ \frac{J}{2N}\sum_{i\neq j} (1- \cos(\theta_i-\theta_j))\,,
\label{Hmf}
\eeq
where $\theta_i$ are angle variables that in our case would correspond to $\theta_i=kx_i$. The analogy becomes explicit in Eq. \eqref{Hmf_self} by using 
$$\Theta^2=\sum_{i,j} \Big( \cos(k(x_i + x_j))+\cos(k(x_i - x_j)) \Big)/(2N^2)\,.$$
Like Hamiltonian $H_{\rm MF}$, also Hamiltonian $H$ is extensive as it satisfies Kac prescription \cite{Campa:2009} for the thermodynamic limit we choose, which keeps $\bar n$ fixed for $N\to\infty$ (see next section). In a canonical ensemble, for $J>0$ the HMF exhibits a second-order phase transition from a paramagnetic to a ferromagnetic phase controlled by the temperature, where the order parameter is the magnetization $M=(M_x,M_y)$ with $M_x=\sum_j\cos\theta_j/N$ and $M_y=\sum_j\sin\theta_j/N$. This suggests to identify $\Theta$ with the $x$-component of a two-dimensional magnetization, and let expect a transition to order for negative values of the detunings, $\Delta_c<0$, for which a non-vanishing interaction potential term tends to minimize the energy (we mention that the dynamics for $\Delta_c>0$ has been recently studied in Ref. \cite{Piazza:2015}). 

Differing from the HMF model, the term $\cos(k(x_i + x_j))$ in $\Theta^2$ originates from the underlying cavity standing-wave potential that breaks continuous translational invariance. Moreover,  the cavity coupling at higher order in $|NU/\Delta_c|$ gives rise to deviations from the Hamiltonian dynamics due to further terms in the LHS of Eq. \eqref{FPE}, which for larger values are responsible for bistable behaviour \cite{Larson:2008} and only in certain limits can be cast in the form of conservative forces. 

We further highlight that long-range correlations can also be established by the terms on the RHS of the FPE in Eq. \eqref{FPE}, which are usually associated with incoherent processes. In fact, retardation effects in the scattering of one atom modify the intracavity potential which traps the whole atomic ensemble. Photon losses, in addition, give rise to sudden quenches of the global potential  \cite{Domokos:2003,Murr:2006}.  When the density is uniform, the terms in the RHS can be reduced to a form \cite{Schuetz:2013} which is analogous to the Brownian Mean Field model \cite{Chavanis}. However, this mapping applies only when the system is deep in the paramagnetic phase. When the atoms form a Bragg grating, instead, damping and diffusion become smaller being the atoms localized at the points where $\sin(kx_j)\sim 0$. Moreover, when  several atoms are trapped in a Bragg grating, also damping and diffusion of atoms which are away from the nodes become smaller. These properties share some analogies with  models constructed to simulate correlated damping \cite{Olmos:2012} and suggest that incoherent dynamics can endorse coherent effects for transient but long times \cite{Schuetz:2014,Schuetz:2015:2}. 

\section{Properties at equilibrium}
\label{Sec:Steady}

We now discuss the existence and the form of the stationary state, namely, of the solution of Eq. \eqref{FPE} satisfying 
$$\partial_tf_{\rm S}=0\,.$$
It is simple to verify that the function of the form 
\beq 
\label{eq:Steady}
f_{\rm S}= f_0 \exp(- \beta H)\,,
\eeq
is a stationary solution in zero order in the parameter $UN/\kappa$ and $\bar \eta$, where $f_0$ warrants normalization. Equation \eqref{eq:Steady} describes a thermal state whose temperature $T$ is solely controlled by the detuning $\Delta_c$:
\beq
\label{Temperature}
k_BT=1/\beta=\frac{\hbar(\Delta_c^2 + \kappa^2)}{-4\Delta_c}\,.
\eeq
We mention that this result has been reported in Ref. \cite{Schuetz:2014}, and was also found in Refs. \cite{Asboth:2005,Niedenzu:2012,Piazza:2014} using different theoretical approaches.

In this section, starting from Eq. \eqref{eq:Steady} we analyse the properties of the system at steady state. We show that Eq. \eqref{eq:Steady} allows to identify the transition to selforganization and the corresponding critical value at which it occurs. By deriving the single-particle free energy in an appropriate thermodynamic limit we demonstrate that the transition to selforganization is a second-order phase transition, whose order parameter is $\Theta$.
We point out that the treatment here presented applies concepts of equilibrium thermodynamics and is strictly valid at the steady state, because it is a thermal distribution.

This section contains analytical results, extracted from Eq. \eqref{eq:Steady}, and data of numerical simulations, obtained by integrating the Stochastic Differential Equations (SDE) which simulate the dynamics of Eq. \eqref{FPE}. These equations have been reported in Ref. \cite{Schuetz:2013} and for completeness are also detailed in Appendix \ref{App:SDE}. 
A single trajectory for $N$ atoms corresponds to integrating the set of coupled equations \eqref{SDE:1} and \eqref{SDE:2} for the variables  $\{x_\ell(t); p_\ell(t)\}$ with $\ell=1,\ldots,N$ and for a given initial condition. From this calculation, for instance, we find $$\Theta(t)=\sum_{\ell=1}^N\cos(kx_{\ell}(t))/N\,.$$
The mean values are numerically computed by taking the average over $n$ such trajectories, which statistically satisfy the initial conditions, and deliver quantities such as $\langle\Theta^2\rangle_t=\sum_{i=1}^n\Theta_i(t)^2/n$, where $i$ now labels the trajectory, $i=1,\ldots,n$.

In the simulations we assume an ensemble of $^{85}$Rb atoms with transition wavelength $\lambda = 780\,\text{nm}$ (D$_2$-line). This gives the recoil frequency $\omega_r = 2\pi \times 3.86\,\text{kHz}$. The transition linewidth is  $\gamma = 2 \pi \times 6\, \text{MHz}$ and the linewidth of the resonator is $\kappa = 2\pi \times 1.5 $ MHz. These parameters correspond to the ones of the experiment of Ref. \cite{Baumann:2010}, they warrant the validity of our semiclassical treatment based on a time-scale separation.

\subsection{Selforganization as second-order phase transition}

In order to characterize the thermodynamic properties of the selforganization transition, we first determine the free energy per particle. Our starting point is the definition of the free energy $F=-k_B T\log \mathcal Z$, where $\mathcal Z$ is the partition function, 
\beq
\label{Z:0}
\mathcal Z=\frac{1}{\Delta^N}\int_{\bm{x}} {\rm d}\bm{x}\int_{\bm{p}}{\rm d}\bm{p}\exp(-\beta H)\,,
\eeq
and $\Delta$ is the unit phase space volume. For convenience, we have introduced the notation $\int_{\bm{x}} {\rm d}\bm{x}\equiv\int_0^\lambda{\rm d}x_1\ldots\int_0^\lambda{\rm d}x_N$ and $\int_{\bm{p}}{\rm d}\bm{p}\equiv\int_{-\infty}^{\infty}{\rm d}p_1\ldots\int_{-\infty}^{\infty}{\rm d}p_N$. After integrating out the momentum variables, Eq. \eqref{Z:0} can be cast in the form
\begin{eqnarray}
\label{eq:partition}
\mathcal{Z}= (Z_0\lambda / \Delta)^N \int_{-1}^{1}{\rm d}\Theta\,\varOmega(\Theta)\exp\left(-N\beta\hbar \bar n\Delta_c\Theta^2\right)\,.
\end{eqnarray}
Here, $Z_0 = ( 2 \pi m / \beta)^{1/2}$ is a constant which depends on the temperature. The functional $\varOmega(\Theta)$ is the density of states at a given magnetization $\Theta$ and is defined as
\beq
\label{Omega}
\varOmega(\Theta)=
\int_{\bm{x}} \frac{{\rm d}\bm{x} }{\lambda^N}\,
\delta\left(\Theta- \frac{1}{N} \sum_{i=1}^{N}\cos(kx_i)\right)\,.
\eeq
For identifying the transition to order, we consider $N\gg 1$. This requires an adequate thermodynamic limit. We choose a thermodynamic limit for which the amplitude $\bar n$, Eq. \eqref{bar:n}, remains constant as $N$ increases and warrants that Hamiltonian in Eq. \eqref{Hmf_self} is extensive. In detail, it corresponds to scale the vacuum Rabi frequency as $g\sim 1/\sqrt{N}$, which is physically equivalent to scale up the cavity mode volume $V$ linearly with $N$, being the vacuum Rabi frequency $g\propto 1/\sqrt{V}$. It follows that the scattering rates characterizing the dynamics scale as $S \sim 1/\sqrt{N}$ and $U\sim 1/N$ as $N\to\infty$ (moreover, $S^2\eta_0 \sim 1/N$, but this contribution is here neglected). Such scaling has been applied in a series of theoretical works \cite{Asboth:2005,Larson:2008,Fernandez-Vidal:2010}. 

With this definition in mind, we determine an explicit form of the free energy as a function of $\Theta$ by using the method of the steepest descent. We identify the fixed point $\Theta^* $, which is given by the equation
\begin{align}
\Theta^* = \frac{I_1 (y\Theta^*) }{ I_0 (y \Theta^*) } \,,
\label{eq:fix-point}
\end{align}
with $y=2\bar{n}/\bar{n}_c$ and $\bar n_c>0$, while $I_1$ and $I_0$ are modified Bessel functions of the first kind \cite{Abramowitz-Stegun} (The details of the calculations are reported in Appendix \ref{App:FreeEnergy}). Depending on $y$, and thus on $\bar n$, Eq. \eqref{eq:fix-point} allows for either one or three solutions, where the two regimes are separated by the value $\bar n=\bar n_c$, with 
\beq
\label{bar:n:c}
\bar n_c=\frac{\kappa^2+\Delta_c^2}{4\Delta_c^2}\,.
\eeq
Using this result, the free energy per particle in the thermodynamic limit takes the form
\begin{align}
\label{F:Landau}
 \mathcal{F}(\Theta)\approx
 \mathcal{F}_0 +
\frac{1}{\beta}\left[\left(1-\frac{\bar n}{\bar n_c}\right)\Theta^2+\frac{1}{4}\Theta^4\right]\,,
\end{align}
with $\mathcal{F}_0 = - k_B T \log (Z_0 \lambda / \Delta )$. Equation \eqref{F:Landau} has the form of the Landau free energy \cite{Landau}, and shows that the transition to selforganization is continuous and of second order. Its form close to threshold for different values of the pump strength, and thus of $\bar n$,  is sketched in Fig. \ref{Fig:1}(d), where $(\Omega / \Omega_c)^2 = \bar{n} / \bar{n}_c$. For $\bar n<\bar n_c$, thus, the order parameter vanishes: The atoms are uniformly distributed in space and one can denote this phase as paramagnetic invoking the analogy between $\Theta$ and a magnetization. For $\bar n>\bar n_c$, on the contrary  the order parameter takes a value different from zero, as shown in Fig. \ref{Fig:1}(c). By setting the first derivative of the free energy, Eq. \eqref{F:Landau}, to zero we also find an analytic expression for the order parameter above but close to the threshold: $\Theta=\pm \sqrt{2(\bar{n}/ \bar{n}_c-1)}$. 

We remark that in Ref. \cite{Asboth:2005} it was conjectured that selforganization in a standing wave cavity is a second-order phase transition. In this section we have demonstrated that this conjecture is correct by performing an explicit mapping of the free energy into the form of a Landau model \cite{Landau}. Our theoretical model demonstrates that the steady state distribution is thermal, it further naturally delivers the steady state temperature and the value of the critical pump strength, here cast in terms of the quantity $\bar n_c$. We observe that the critical value $\bar n_c$ is in agreement with the value determined in Ref. \cite{Asboth:2005} by means of a mean field model based on a phenomenological derivation (This is visible after considering the definition in Eq. \eqref{bar:n}, which gives the critical pump strength value $\Omega_c$ after using $S_c=g\Omega_c/\Delta_a$ as a function of the critical value $\bar n_c$ of Eq. \eqref{bar:n:c}). In Ref. \cite{Niedenzu:2012} the self-organization threshold was estimated by means of a kinetic theory based on treating the
cavity field semiclassically, finding a value consistent with our result. 

We remark that, the typical concept in second-order phase transition of spatial domains, whose average size increases with a power-law behaviour as the critical value is approached, becomes now invalid: Their energetic cost scales with the system size due to the long-range cavity-mediated potential. This is simply understood as two domains with $\langle \Theta\rangle=+1$ and $\langle \Theta\rangle=-1$ generate fields which interfere destructively, resulting in a vanishing intracavity photon number. This example illustrates the non-additivity of long-range interacting systems. We now analyse more in detail the behaviour of the magnetization.

\subsection{Phase diagram}

The magnetization of our model, Eq. \eqref{Theta}, is intrinsically related to the spatial order of the atoms within the cavity, and thus determines the properties of the signal at the cavity output. Its stationary value depends on the various physical quantities, which can be summarized in terms of the single parameter $\bar n$ in Eq. \eqref{bar:n}. The detuning $\Delta_c$, which also enters in the definition of $\bar n_c$, determines the temperature of the steady state, see Eq. \eqref{Temperature}. 

\begin{figure}[hbtp]
\begin{center}
\includegraphics[width=0.48\textwidth]{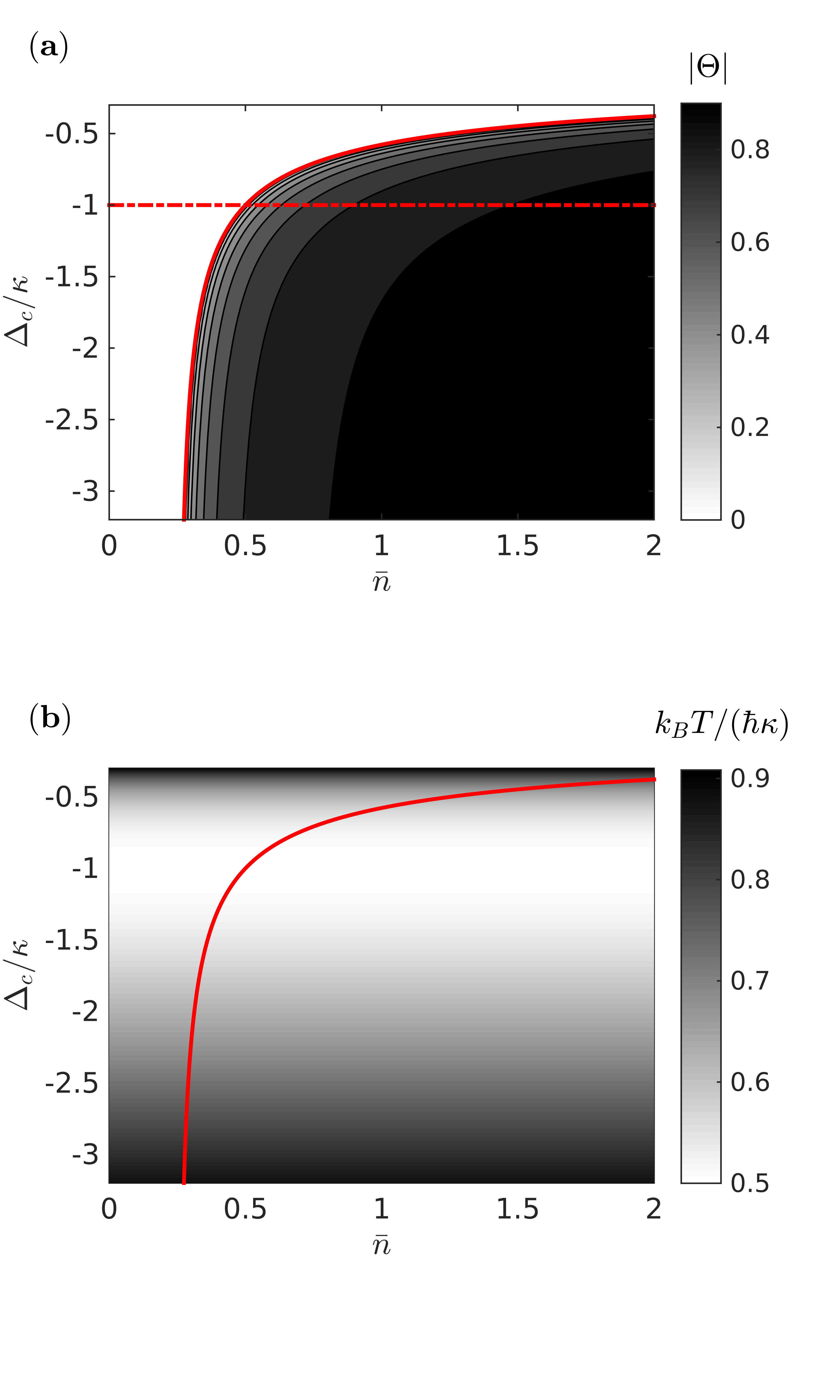} 
\caption{(color online) (a) Order parameter $|\Theta|$ and (b) steady-state temperature as a function of $\bar n$ and $\Delta_c$ (in units of $\kappa$). The red line denotes the value $\bar n_c$ as a function of $\Delta_c$, as reported in Eq. \eqref{bar:n:c}. 
 }
\label{Fig:PD}
\end{center}
\end{figure}

Figure  \ref{Fig:PD}(a) displays the phase diagram of the magnetization as a function of $\bar n$ and $\Delta_c$: the white region is the paramagnetic phase, the dark region the ferromagnetic one, while the scale of grey indicates the value of $|\Theta|$. We note that the lines at constant $\Delta_c$ correspond to constant asymptotic temperatures and to a well defined threshold value of $\bar n_c(\Delta_c)$. Following one such line, the value of $|\Theta|$ is zero for $\bar n<\bar n_c$, while above $\bar n_c$ it grows monotonically till unit as $\bar n\to \infty$. The magnetization as a function of $\bar n$ and at $\Delta_c=-\kappa$ is shown in Fig. \ref{Fig:1}(c). 

Keeping $\bar n$ fixed and varying $\Delta_c$, instead, consists in varying the temperature. However, not for all values of $\bar n$ there exists a temperature at which the transition to ferromagnetism is observed. In fact, if $\bar n<{\rm min}(\bar n_c)=1/4$, the phase is paramagnetic for all values of $\Delta_c$.  For $\bar n>1/4$, instead, there exists a critical value of $\Delta_c(\bar n)$ at which the transition to selforganization occurs. In this case, above threshold the magnetization monotonically grows with $\Delta_c$. The temperature of the atoms is shown in Fig. \ref{Fig:PD}(b): here it is clearly visible that the temperature is independent on $\bar n$ and is solely a function of $\Delta_c$. In particular, it reaches a minimum at $\Delta_c=-\kappa$, as one can verify using Eq. \eqref{Temperature}. The corresponding minimal temperature is $k_BT_{\rm min}=\hbar\kappa/2$. 

\subsection{Dynamics of the magnetization at steady state}

The mapping of the free energy to Landau model allows one to draw an analogy between selforganization and ferromagnetism. Due to the long-range interactions, however, the symmetry-breaking transition does not occur through the spatial formation of magnetized domains of increasing size, rather through the observation of Bragg gratings during long period of times, whose mean duration increases as the pump strength is increased above threshold. This property was already reported in Refs. \cite{Domokos:2002,Asboth:2005} and is also found in the HMF \cite{Campa:2009}. The behaviour close to threshold is instead to large extent unexplored, as it is characterized by large fluctuations of the cavity field and thus requires a theoretical model that treats the cavity field as a quantum variable, what our model does. Our analysis focuses on the statistical properties of these time intervals, and more generally of the autocorrelation function of the magnetization across the transition. In this section we discuss this temporal behaviour by analysing trajectories of the magnetization evaluated by means of the SDE as in Appendix \ref{App:SDE}. We set $\Delta_c = - \kappa$ and $N |U| / \kappa = 0.05$.

\subsubsection{Stationary magnetization for finite $N$.}

\begin{figure}[hbtp]
	\begin{center}
		\includegraphics[width=0.47\textwidth]{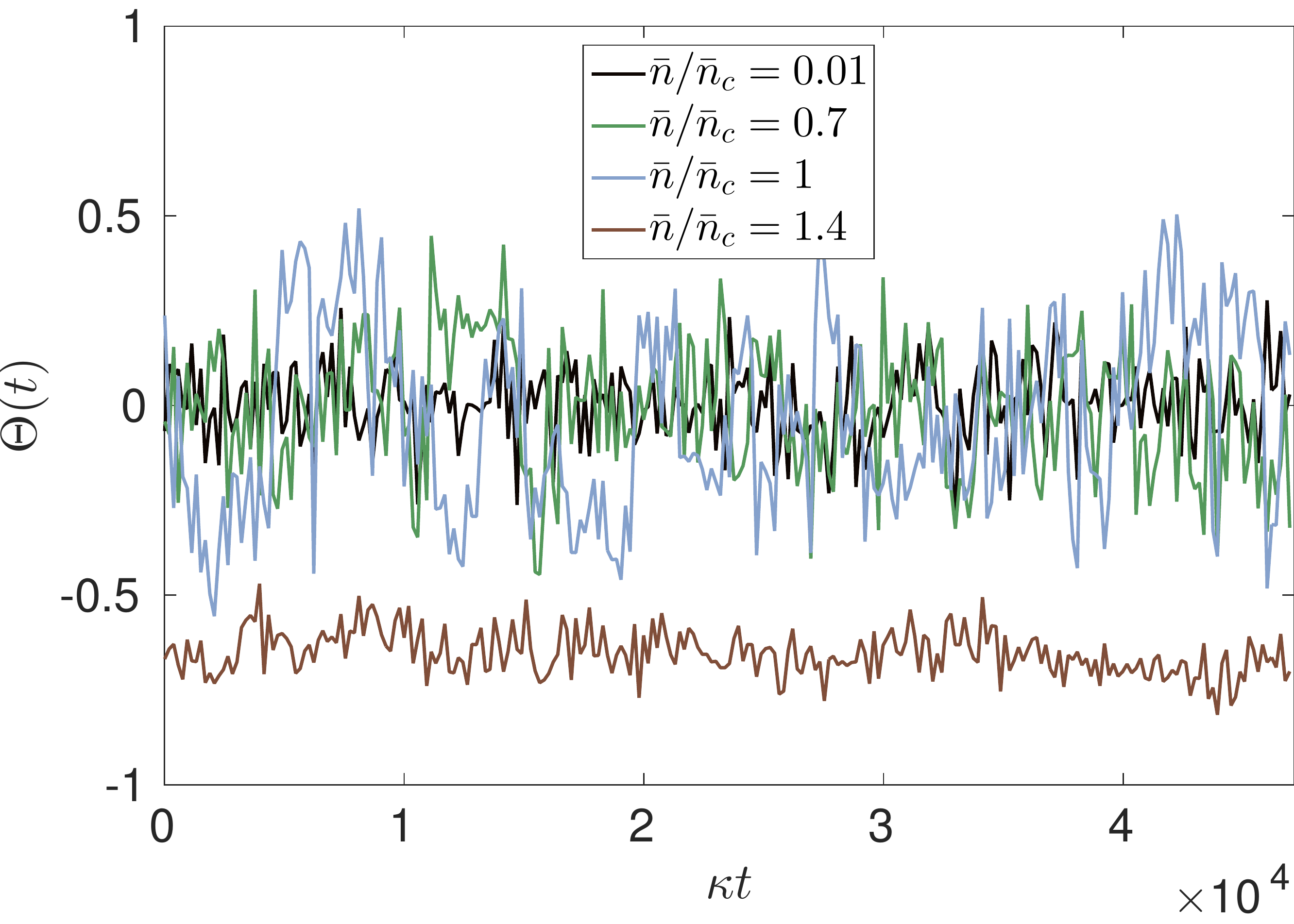} 
		\caption{(color online) Order parameter as a function of time (in units of $\kappa^{-1}$) at the asymptotics of the dynamics and for different values of $\bar n$ (see inset). Each trajectory corresponds to a numerical simulation with $N=50$ atoms.}
		\label{fig:trajectories}
	\end{center}
\end{figure}

In order to perform the numerical analysis, we first benchmark the statistical properties for a finite number of trajectories. Typical trajectories at the steady state are shown in Fig. \ref{fig:trajectories} for different values of $\bar n$. 
\FloatBarrier
They show $\Theta(t)$, obtained by averaging over the instantaneous positions of 50 atoms within the resonator. Fluctuations about the mean value are visible: their size increases below threshold as $\bar n$ is increased and depends on the number of atoms, as one can see in Fig. \ref{fig:Ptheta} (see below). In order to extract the order parameter from the numerical data we thus need to estimate the size of the fluctuations about the mean value as a function of $N$. For this purpose we determine the probability distribution $P_N(\Theta_0)$ of finding $\Theta=\Theta_0$ at the stationary state, which we define as
\beq
P_N(\Theta_0)= 
\mathcal P_0 \int_{-1}^{1}{\rm d}\Theta\,\delta(\Theta-\Theta_0) \varOmega(\Theta)\exp\left(-\beta \hbar \Delta_c \bar n N \Theta^2\right) \,,
\label{eq:PTheta_0}
\eeq
where $\varOmega(\Theta)$ is given in Eq. \eqref{Omega} and the parameter $\mathcal P_0=(Z_0\lambda/ \Delta)^N/\mathcal{Z}$ warrants normalization: $\int_{-1}^1 {\rm d} \Theta_0 P_N (\Theta_0) = 1$. For a given detuning $\Delta_c$ this probability distribution depends on $\bar n$ and on the atom number $N$. We determine $P_N(\Theta_0)$ using our analytical model and performing the integral by means of the Metropolis algorithm \cite{Metropolis}. 

\begin{widetext}
	\begin{figure*}[t]
		\includegraphics[width=1\textwidth]{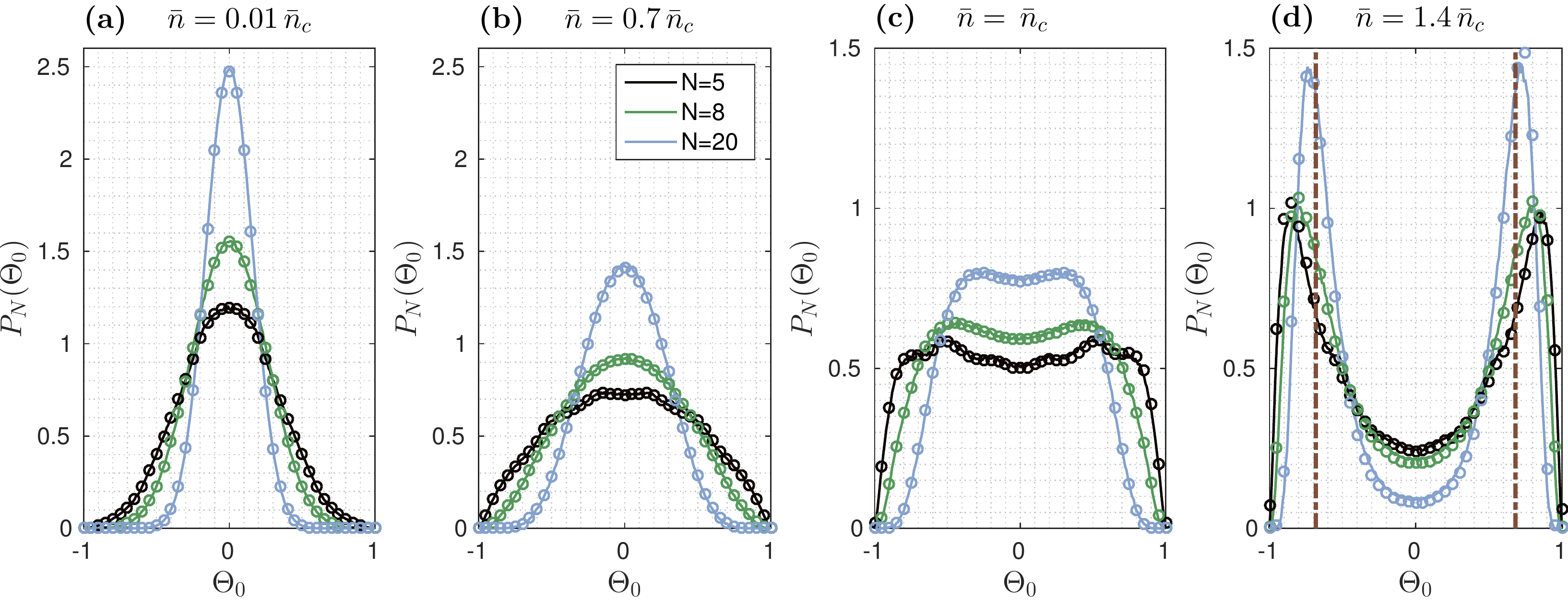}
		\caption{(color online) Probability distribution for the order parameter at steady state, $P_N (\Theta_0)$ as in Eq. \eqref{eq:PTheta_0}, for $N=5,8,20$ atoms  with $\Delta_c = - \kappa$ and $\bar n/\bar n_c=0.01,\,0.7,\,1,\,1.4$ (from left to right). The dots correspond to the probability distribution $P_N(\Theta_0)$ extracted from numerical simulations at steady state, performed by means of the SDE. The dashed vertical lines in (d) indicate the asymptotic value $\Theta_0 = \pm \Theta^*$, Eq. \eqref{eq:fix-point}, for $\bar{n} = 1.4 \, \bar{n}_c$.}
		\label{fig:Ptheta}
	\end{figure*}
\end{widetext}

The results are displayed in Fig. \ref{fig:Ptheta}  for different atom numbers $N$ and pumping strengths $\bar n$. The curves clearly show that the size of the fluctuations about the mean value decrease with $N$. We also observe that, for $N$ fixed, the fluctuations about the mean value increase with $\bar n$ as it approaches the threshold value from below. For atom numbers of the order of 50 and larger we verified that $P_N (\Theta_0)$ converges to the form $\exp(-N \,\Theta_0^4/4)$ for $\bar{n} = \bar n_c$, in agreement with the result found in the thermodynamic limit. Above threshold, on the contrary, the distribution exhibits two peaks whose centers converge towards the asymptotic values of Eq. \eqref{eq:fix-point} for large $N$ and whose widths decrease as $\bar n$ is increased. We compare these results with the data obtained after integrating the SDE (circles), and verify the convergence of the numerical results with increasing $N$ to the predictions at the thermodynamic limit. 

\begin{figure}[hbtp]
\begin{center}
\includegraphics[width=0.48\textwidth]{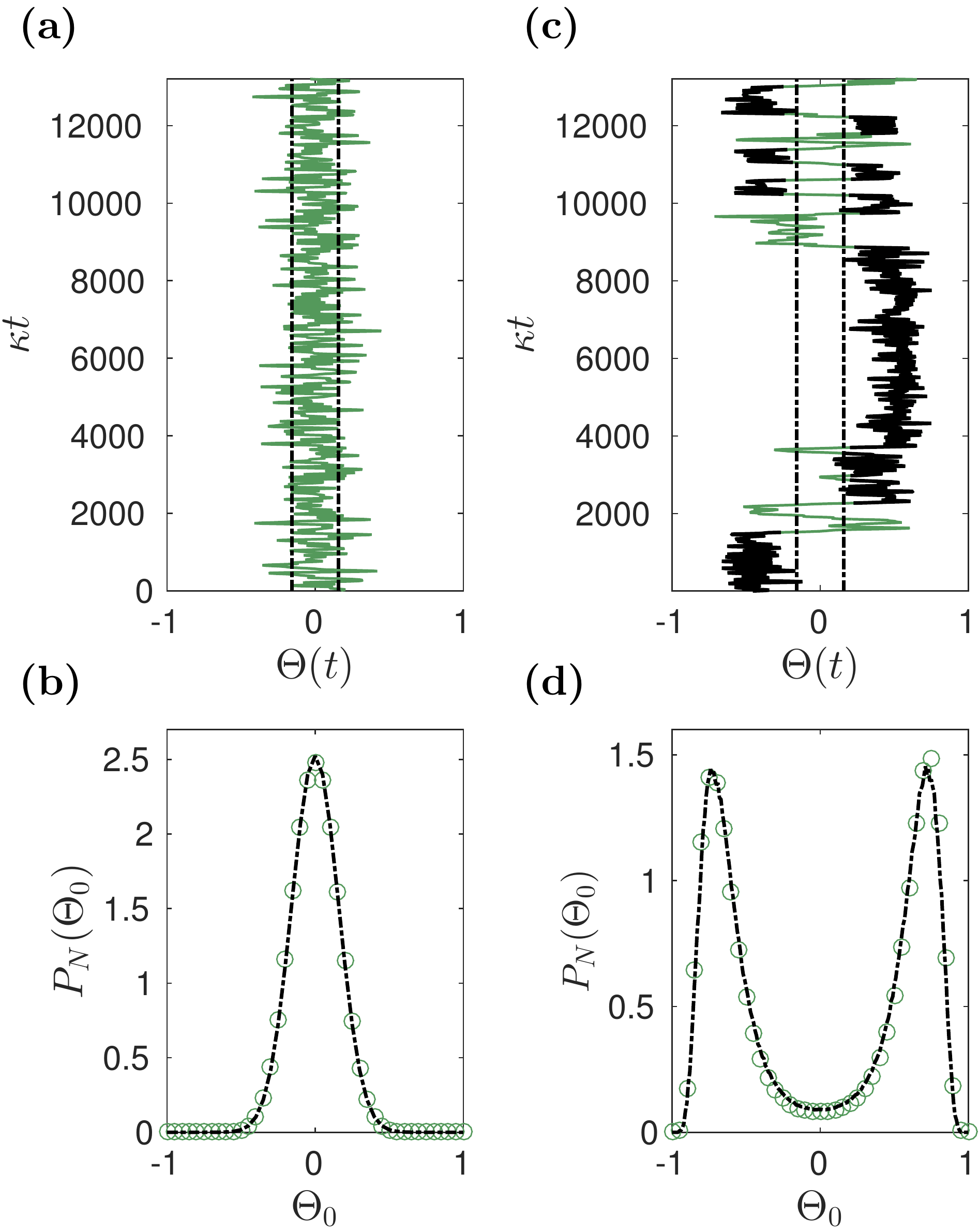}
\caption{(color online) Upper panels: Magnetization $\Theta$ as a function of time (in units of $\kappa^{-1}$), obtained from a simulation of the SDE for $N=20$, $\Delta_c=-\kappa$ and $\bar{n} = 0.01 \,\bar{n}_c$ (a),  $\bar{n} = 1.4 \,\bar{n}_c$ (c). The black dashed lines are located at $\pm \sigma_N = \pm \sqrt{1/(2N)}$ and indicate the statistical uncertainty in the determination of the value of $\Theta_0$. Subplots (b) and (d) display the corresponding probability distribution $P_N(\Theta_0)$ obtained after averaging over time and over 100 trajectories $\Theta(t)$ (circles). The dashed line in (b) is the theoretical prediction in Eq. \eqref{P:theo}. The dashed line in (d) corresponds to the distribution obtained by numerically integrating Eq. \eqref{eq:PTheta_0} using a Metropolis algorithm \cite{Metropolis}.}
\label{fig:dyn}
\end{center}
\end{figure}

Figure \ref{fig:dyn}(a) displays $\Theta(t)$ as a function of time obtained by integrating the SDE for $N=20$ atoms and $\bar n = 0.01 \,\bar{n}_c$, thus well below threshold. The distribution $P_N(\Theta_0)$ that we extract after averaging over the time and over 100 trajectories of this sort is given by the circles in Fig. \ref{fig:dyn}(b). The curve is in excellent agreement with a Gaussian distribution centered at $\Theta_0=0$ (dashed curve) whose explicit derivation is reported in Appendix \ref{App:Estimate} and which reads 
\beq
\label{P:theo}
P_N^{\rm theo}(\Theta_0) = \frac{1}{\sqrt{2 \pi \sigma_N^2}} \exp\left(-\frac{\Theta_0^2}{2 \sigma_N^2}\right)\,,
\eeq
with 
\beq
\sigma_N=1/\sqrt{2N}\,.
\eeq
From this result we identify the width $\sigma_N$ with the statistical uncertainty in determining the value of $\Theta_0$.
Figure \ref{fig:dyn}(c) displays a trajectory $\Theta(t)$ for $\bar n =1.4\, \bar n_c$, thus above threshold; the corresponding distribution $P_N(\Theta_0)$ is given by the circles in Fig. \ref{fig:dyn}(d). The trajectory exhibits jumps between the two values of the Bragg gratings, the duration of the time intervals during which the atoms are trapped in a Bragg grating determines the size of the fluctuations about the two peaks of the probability distribution, the finite rate at which these jumps occur is the reason for the non-vanishing value of the probability at $\Theta_0\sim 0$. \\

\subsubsection{Autocorrelation function.}

\begin{figure}[hbtp]
\begin{center}
\includegraphics[width=0.48\textwidth]{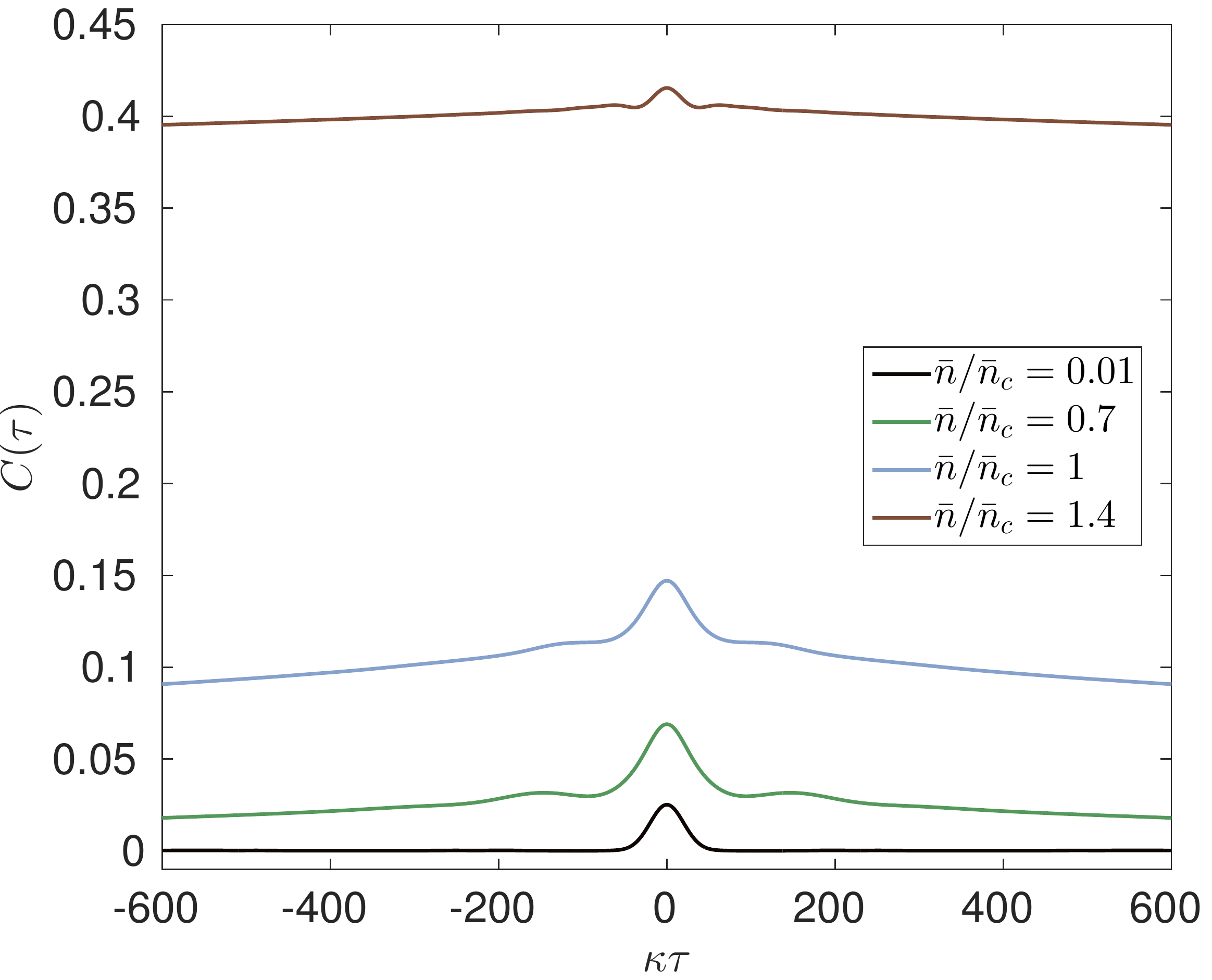}
\caption{(color online) Autocorrelation function $C(\tau) = \lim_{t\to\infty} \langle \Theta(t) \Theta(t+\tau) \rangle$, Eq. \eqref{eq:autocorr}, as a function of the time $\tau$ (in units of $\kappa^{-1}$) for $N=20$ atoms, $\Delta_c=-\kappa$,  and various values of $\bar{n}$ (see inset). The curves are obtained by determining $\Theta(t)$ with the numerical data (SDE).}
\label{auto:corr}
\end{center}
\end{figure}

We now analyse the autocorrelation function for the magnetization,
\beq
\label{eq:autocorr}
C(\tau) = \lim_{t\to\infty} \langle \Theta(t) \Theta(t + \tau) \rangle\,,
\eeq 
which we extract from the trajectories evaluated using the SDE. Figure \ref{auto:corr} displays $C(\tau)$ for different values of $\bar n$. For all values of the pump strength a fast decaying component is always present, whose temporal width seems to be independent of $\bar n$. One also notices the contribution of a slowly decaying component, whose decay rate decreases as $\bar n$ increases. 

\begin{figure}[hbtp]
\begin{center}
\includegraphics[width=0.48\textwidth]{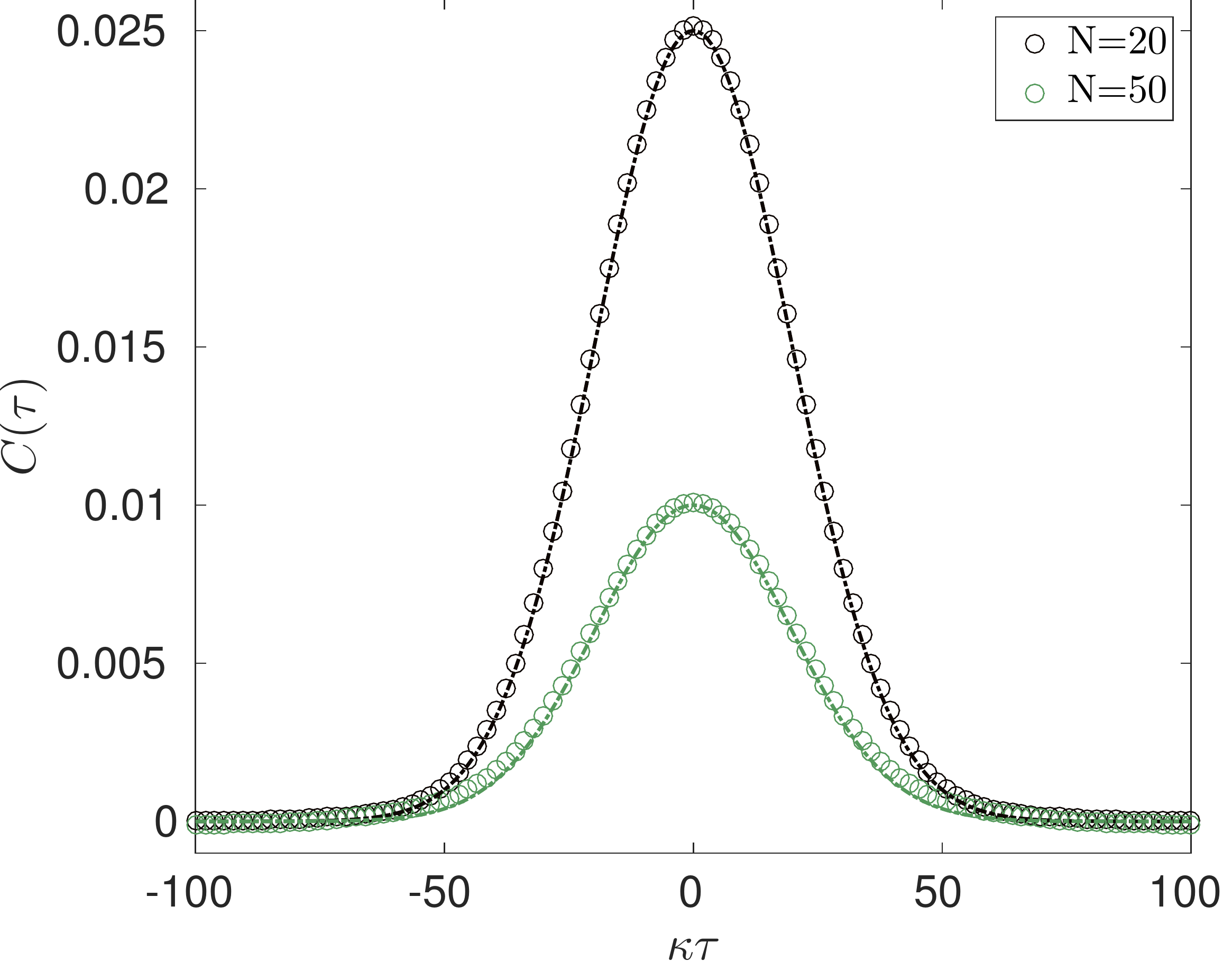}
\caption{(color online) Autocorrelation function $C(\tau) = \lim_{t\to\infty} \langle \Theta(t) \Theta(t+\tau) \rangle$ as a function of the time $\tau$  (in units of $\kappa^{-1}$) for $N=20$ and $N=50$ atoms (see inset). The circles correspond to numerical simulations performed  with $\bar{n} = 0.01\,\bar{n}_c$ and $\Delta_c=-\kappa$. The line shows the analytical estimate using Eq. \eqref{eq:corr:theo}.}
\label{fig:corrbelow}
\end{center}
\end{figure}

In order to gain insight, we first analyse the autocorrelation function below threshold, for $\bar{n} = 0.01\, \bar{n}_c$. For this case we can reproduce the numerical result by means of an analytical model, reported in Appendix \ref{App:Estimate}. This model assumes that the atoms are homogeneously distributed in space and form a thermal distribution at the temperature determined by  Eq. \eqref{eq:hbar_beta}, which corresponds to the stationary solution of the FPE in Eq. \eqref{FPE}  well below threshold \cite{Schuetz:2013}. Starting from this state, their motion is assumed to be ballistic, and is thus calculated after setting $\bar n=0$ in Eq. \eqref{FPE}. The resulting autocorrelation function reads 
\beq
\label{eq:corr:theo}
C_{\rm free}(\tau)= \sigma_N^2 \exp\big(- ( \tau / \tau_c^{\rm free} )^2\big)\,,
\eeq 
where the correlation time is
\beq
\tau_c^{\rm free} = \sqrt{ \hbar \beta /  \omega_r}\,.
\eeq
Its excellent agreement with the numerics is visible in Fig. \ref{fig:corrbelow}. This result shows that below threshold the fluctuations are mostly due to thermal motion, while the effect of the cavity forces, which tend to localize the atoms, is negligible. By considering the analogy between the different curves in Fig. \ref{auto:corr}, we conjecture that thermal fluctuations are responsible for the short-time behaviour of the autocorrelation function. 

We now turn to the long-time behaviour of the autocorrelation function for increasing values of $\bar n$. Inspection of typical trajectories close and above thresholds, shown in Figs. \ref{fig:trajectories} and \ref{fig:dyn}(c), show that this is related to the time scales over which the atomic ensemble forms a Bragg grating. The system can take on values for the collective parameter $\Theta$ clearly exceeding the value of $\sigma_N$  for times which are orders of magnitude larger than the correlation time $\tau_c$ characteristic of thermal fluctuations, as visible in Fig. \ref{fig:dyn}(c). We call these finite time intervals {\it trapping times}, corresponding to configurations in which (part of) the atoms are trapped in Bragg gratings. 

\begin{figure}[hbtp]
\begin{center}
\includegraphics[width=0.48\textwidth]%
{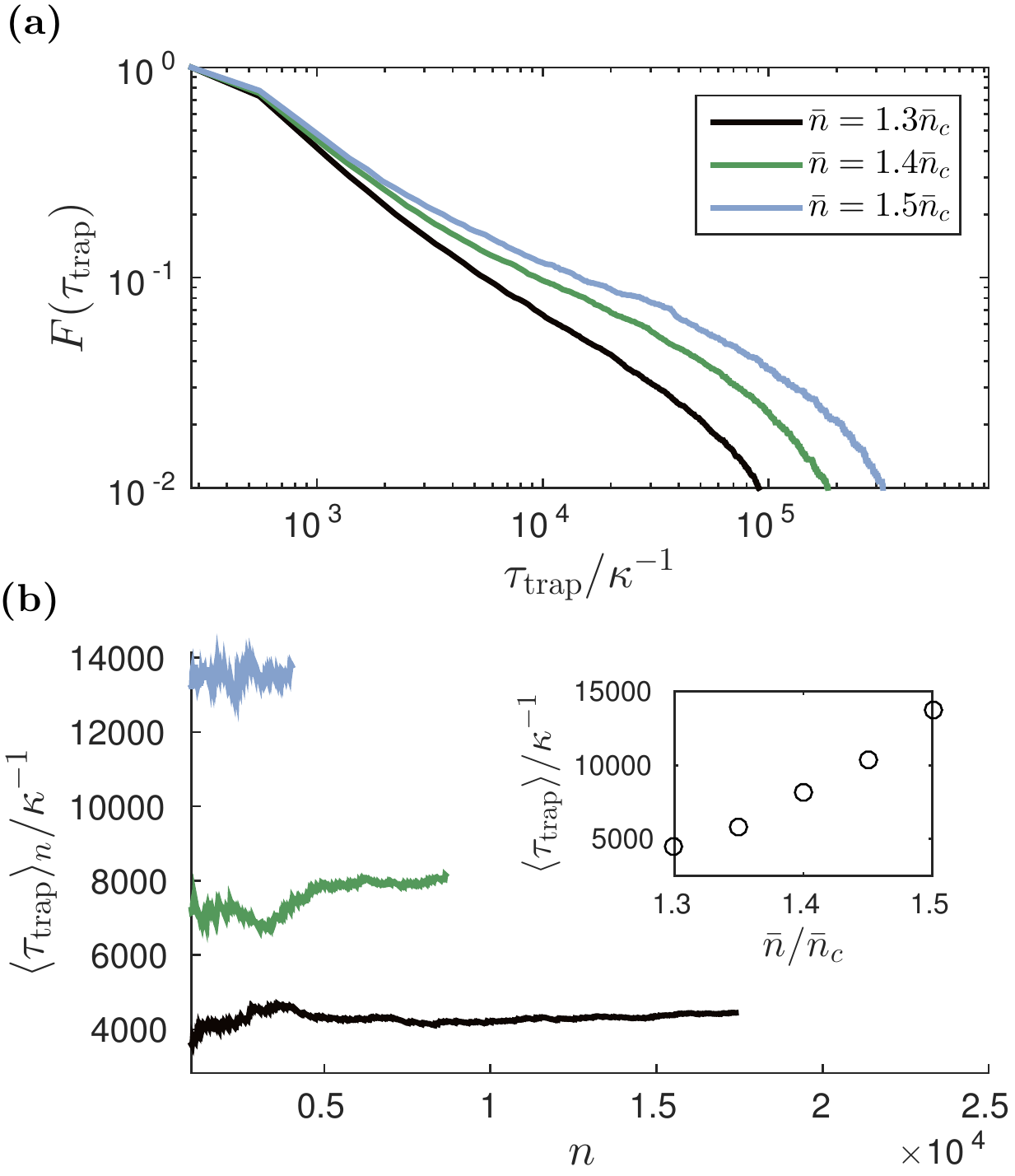}
\caption{(color online) Statistics of the trapping times, evaluated numerically by averaging over 100 trajectories of $N=20$,  $\Delta_c=-\kappa$, total evolution time $t_{\rm tot}\approx 10^6\kappa^{-1}$. The curves correspond to  different values of $\bar n$ above threshold (see inset).  (a) Cumulative distribution $F(\tau_{\text{trap}})$ for the trapping times, Eq. \eqref{F:tau}. Higher pumping strengths lead to longer trapping times. Subplot (b) displays the mean trapping time $\langle \tau_{\text{trap}} \rangle_n$, Eq. \eqref{eq:trapped_n}, as a function of the number of counts $n$. The inset shows the values of $\langle \tau_{\text{trap}} \rangle$ as a function of $\bar n$  which we extrapolate from the curves, like the ones shown in the onset.}
\label{fig:trapping}
\end{center}
\end{figure}

In order to analyse the statistics of the trapping times, we first introduce the following criterion: the atoms are forming a Bragg grating when  $|\Theta (t)| > \sigma_N$. This criterion alone, however, also includes fluctuations that can also happen well below threshold, as visible in Fig. \ref{fig:dyn} (a). For this reason we set an infrared cutoff for the trapping times, such that they shall exceed $\tau_c^{\rm free}$. Herewith, we thus find a trapping time of length $\tau_{\text{trap}}$ with starting point $t$ and end point $t+\tau_{\text{trap}}$ if  $|\Theta (t+t')| > \sigma_N$ for $ t' \in [0,\tau_{\text{trap}}]$ and $\tau_{\text{trap}} > 10 \, \tau_c^{\rm free}$. It is important to note that this sets a rather strict criterion on the trapping times as we will explain now. In Fig. \ref{fig:dyn} (c), one can see that even if the atoms seem to be trapped in a grating, the order parameter can take on values $|\Theta(t)| < \sigma_N$ for times of the order of $\tau_c^{\rm free}$. We choose to ignore these events when they are not associated with a sign change of $\Theta$. We perform the statistics of the trapping times by evaluating the probability density $P_{\rm trap}(\tau)$ of finding a trapping time of length $\tau$, and then using this quantity to determine the cumulative distribution $F(\tau_{\text{trap}})$, defined as 
\begin{equation}
\label{F:tau}
F(\tau_{\text{trap}})
=\int_{\tau_{\text{trap}}}^{\infty} {\rm d}\tau' P_{\rm trap}(\tau')\,.
\end{equation}
Distribution $F(\tau_{\text{trap}})$ thus gives the probability that the trapping time is larger than $\tau_{\text{trap}}$. Figure \ref{fig:trapping}  displays $F(\tau_{\text{trap}})$, as we extracted it for  $N=20$ atoms and different values of $\bar n$: It is clearly visible that the trapping times are shifted towards higher values as $\bar{n}$ increases. The distribution exhibits long tails, which suggests that this dynamics is characterized by the existence of rare events with very long trapping times. In order to better understand this behaviour, we determine the mean trapping time $\langle \tau_{\text{trap}} \rangle_n$. This is numerically found for a given interval of time $t_{\rm tot}$, in which $n$ trapping intervals of length $\tau_{\text{trap}}^{(i)}$ are counted ($i=1,\ldots,n$), and reads
\begin{align}
\langle \tau_{\text{trap}} \rangle_n =  \sum_{i=1}^n \tau_{\text{trap}}^{(i)}/n\,.
\label{eq:trapped_n}
\end{align}
In Fig. \ref{fig:trapping} (b) we plot $\langle \tau_{\text{trap}} \rangle_n$ as a function of the number of counts for $N=20$ and various values of $\bar n$ above threshold. The mean trapping time $\langle \tau_{\text{trap}} \rangle_n$, in particular, seems to converge to a finite value for sufficiently long integration times. We argue, however, that this can be an artifact of the finite integration time $t_{\rm tot}$, which we choose to be $t_{\rm tot} \approx 10^6\kappa^{-1}$: This conjecture is supported by the rather steep decay of the cumulative distribution at $t > 10^5\kappa^{-1}$ visible in Fig. \ref{fig:trapping} (a). Hence, our results do not exclude the existence of a power-law decay of the distribution $F(\tau)$. 
This discussion clearly shows, nevertheless, that the trapping times are responsible for the long tails of the autocorrelation function.

\begin{figure}[hbtp]
\begin{center}
\includegraphics[width=0.48\textwidth]%
{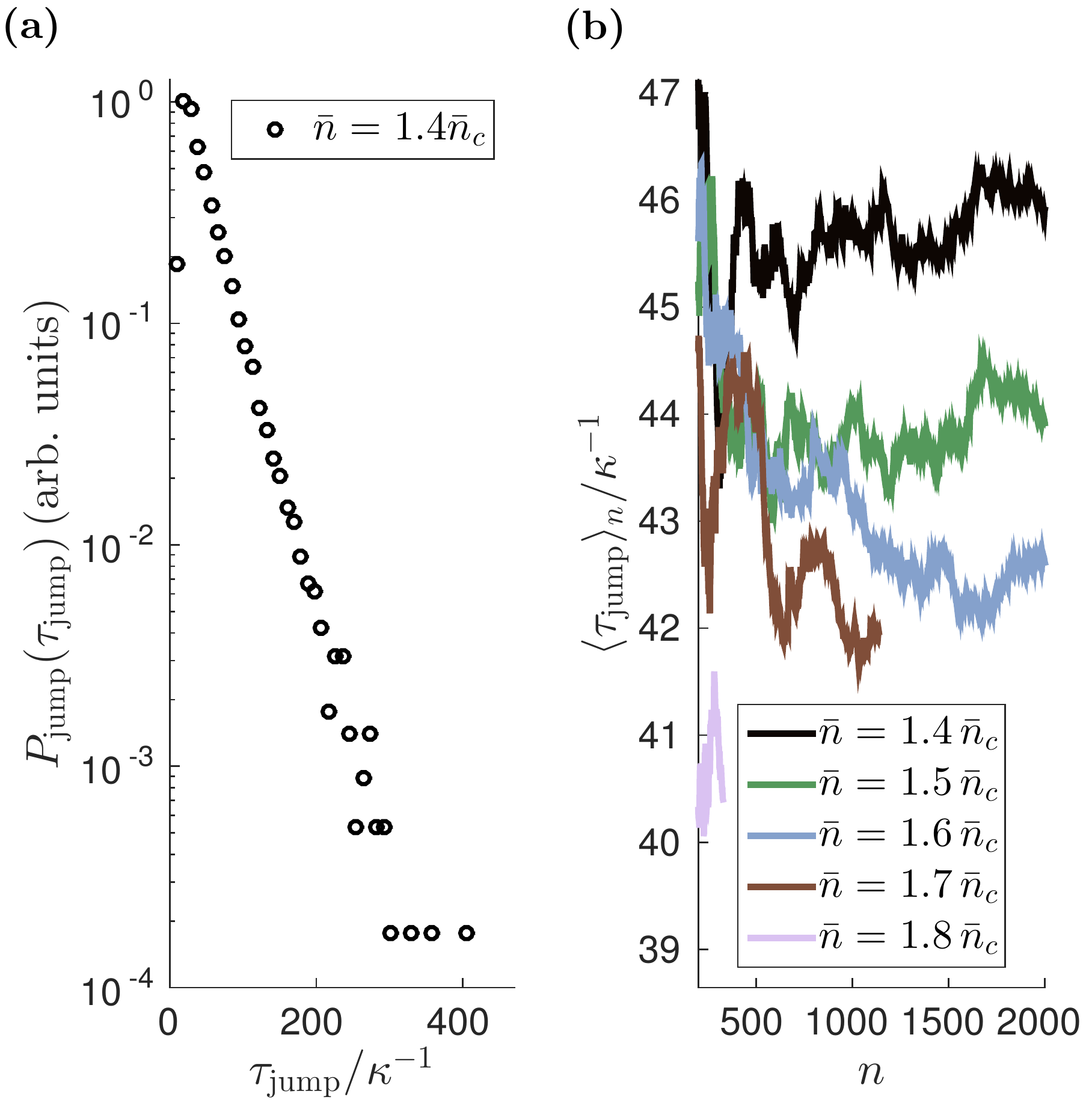}
\caption{ (color online) Statistics of the jumping times, evaluated numerically by averaging over 100 trajectories of $N=20$,  $\Delta_c=-\kappa$, total evolution time $t_{\rm tot}\approx 10^6\kappa^{-1}$.  (a) Probability distribution $P_{\text{jump}} (\tau_{\text{jump}})$ for $\bar{n} = 1.4 \,\bar{n}_c$. (b) Mean jumping time $\langle \tau_{\text{jump}}\rangle_n$, Eq. \eqref{eq:jump_n}, as a function of the number of counts $n$ and for several values of $\bar{n}$ above threshold (see inset). }
\label{fig:jumping}
\end{center}
\end{figure}

We now study the statistics of the events which lead to jumps between two Bragg gratings. These events are visible, for instance, in Fig. \ref{fig:dyn} (c), and are characterized by a time scale which we now analyze. We denote these finite times by {\it jumping times}. More precisely, we define a jump of time length $\tau_{\text{jump}}$ as the interval of time $[0 , \tau_{\text{jump}}]$ within which $|\Theta(t+t')|< \sigma_N$ for $t' \in [0 , \tau_{\text{jump}}]$. We further impose that at the starting and the end points of the jumps the order parameter $\Theta$ has a different sign, such that the configuration has switched, for instance, from an even pattern  $(\Theta > \sigma_N)$ to an odd one $(\Theta < -\sigma_N)$. 
We identify jump events in Fig. \ref{fig:dyn} (c) with the green segments. An exception is the event at $\kappa t \sim 3000$, which does not fulfill the criteria we impose and thus does not qualify.
We numerically determine the probability distribution $P_{\text{jump}}(\tau_{\text{jump}})$ for the jumping times at a given value of $\bar n>\bar n_c$. Figure \ref{fig:jumping} (a) displays the probability distribution $P_{\text{jump}}(\tau_{\text{jump}})$ for  $\bar{n} = 1.4\,\bar{n}_c$. We observe that it exhibits the features of exponential decay with time. Further information is extracted from the mean jumping time $\langle \tau_{\text{jump}} \rangle_n$, which we evaluate as 
\begin{align}
\label{eq:jump_n}
\langle \tau_{\text{jump}} \rangle_n =  \sum_{i=1}^n \tau_{\text{jump}}^{(i)}/n\,,
\end{align} 
with $\tau_{\text{jump}}^{(i)}$ the jumping time for the $i$-th jump and $i=1,\ldots n$. Figure \ref{fig:jumping} (b) displays $\langle \tau_{\text{jump}} \rangle_n$ for different pumping strengths. The mean values $\langle \tau_{\text{jump}} \rangle_n$ do not differ much for different pumping strengths, in agreement with the conjecture that thermal fluctuations are responsible for the short-time behaviour of the autocorrelation function. Nevertheless, we see indications that the mean jumping time decreases as $\bar n$ increases, thus at large pump strengths the atoms reorganize in Bragg gratings over shorter time scales. \\

\begin{figure}[hbtp]
\begin{center}
\includegraphics[width=0.48\textwidth]{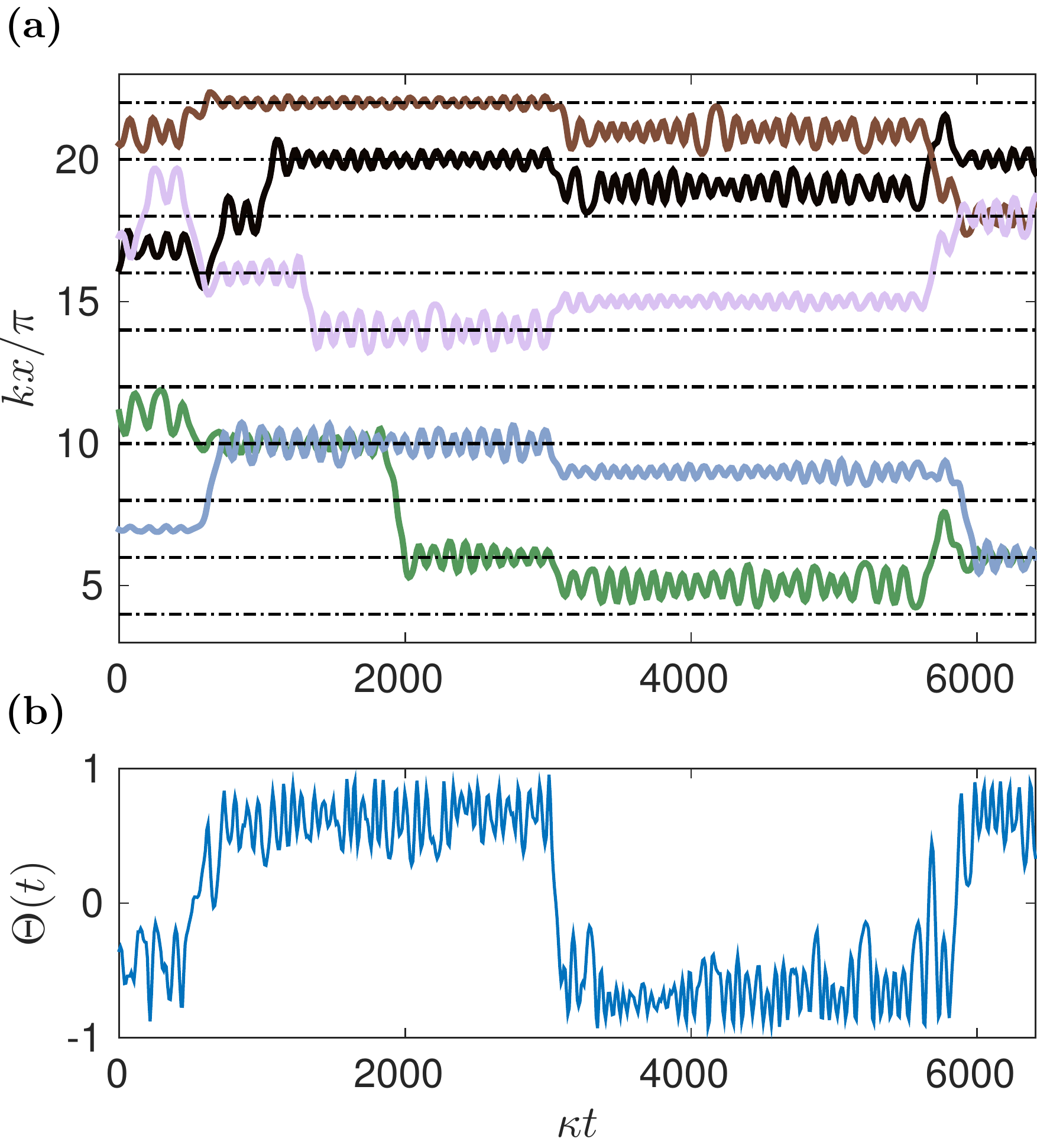}
\caption{ (color online) (a) Individual atomic trajectories and (b) corresponding order parameter as a function of time (in units of $\kappa^{-1}$) for $N=5$ atoms, $\Delta_c=-\kappa$, and $\bar{n} = 1.4 \, \bar{n}_c$. The black dashed horizontal lines in (a) indicate the position of the even sites of the cavity spatial mode function. The trajectories have been numerically evaluated  taking the stationary state as initial distribution.}
\label{fig:singleatoms}
\end{center}
\end{figure}

Insight into the dynamics underlying a jump in the order parameter can be gained by considering the corresponding individual atomic trajectories. A simulation for  $N=5$ atoms is shown in Fig. \ref{fig:singleatoms} (a) for the choice  of a pump strength above threshold $\bar{n} = 1.4 \,\bar{n}_c$. At a given instant of time, the atomic positions  are in general at distances which are integer multiples of the cavity wavelength, thus localized  either at the even or the odd sites of the spatial mode function, thus forming one of the two possible Bragg gratings. When this occurs, the atoms perform oscillations about these positions. The amplitude of these oscillations does not remain constant, and one can observe an effective exchange of mechanical energy among the atoms. 
This can lead to a change of the potential that can untrap atoms. 
The onset of this behaviour seems to be the precursor of the instability of the whole grating, as one can observe by comparing these dynamics with the one of the corresponding order parameter in subplot (b). The oscillations about the grating minima, moreover, are responsible for the damped oscillation observed in the autocorrelation function in Fig. \ref{auto:corr} for values of $\bar n$ above threshold.

\subsubsection{Power spectrum.}

\begin{figure}[hbtp]
\begin{center}
\includegraphics[width=0.48\textwidth]{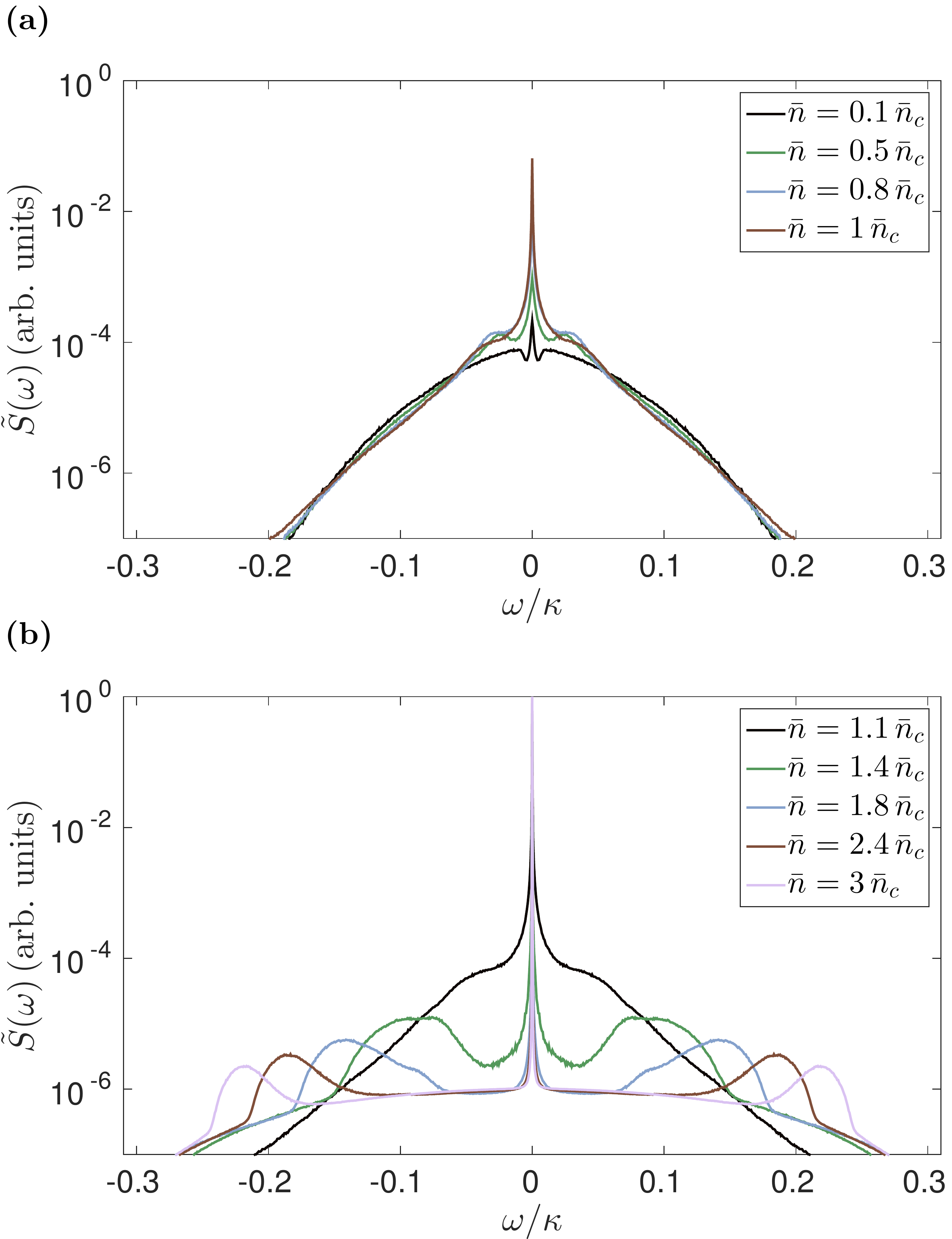}
\caption{(color online) Spectrum of the autocorrelation function $\tilde{S}(\omega)$, Eq. \eqref{S:w} and in arbitrary units, as a function of the frequency (in units of $\kappa$) for different $\bar n$, and evaluated from the numerical data of  $\Theta (t)$  for 100 trajectories of $N=50$ atoms, $\Delta_c=-\kappa$, and evolution time $t_{\rm tot}=10^4\kappa^{-1}$. The subplots show the spectrum for $\bar{n}$ below (a) and above (b) threshold (see insets).}
\label{fig:cuts}
\end{center}
\end{figure}

Complementary information to the temporal behaviour of the autocorrelation function can be gained by studying its Fourier transform. We thus 
numerically compute the power spectrum of $\Theta(t)$, which we define as
\begin{equation}
\label{S:w}
\tilde{S}(\omega) =\langle |\Theta(\omega)|^2\rangle\,,
\end{equation}
where
\begin{equation}
\Theta(\omega)=\int_0^{t} {\rm d}\tau \exp(- {\rm i} \omega \tau) \Theta (\tau) 
\end{equation}
is the Fourier transform of the order parameter. Figure \ref{fig:cuts} displays the spectrum of the autocorrelation function for different values of $\bar n$ (a) below and (b) above threshold. 

One clearly observes two different kinds of behaviour, depending on whether $\bar n$ is  below or above threshold: For $\bar n<\bar n_c$ we observe a rather broad spectrum about $\omega = 0$, whose breadth increases as $\bar n$ approaches the critical value from below. The emergence of a flat broad structure can be associated with the creation of (unstable) Bragg gratings, and is related to the broadening of the distribution $P_N(\Theta_0)$ visible in Fig. \ref{fig:Ptheta}(b)-(c). Above threshold, for $\bar{n} > \bar{n}_c$, the width of the component centered at zero frequency becomes dramatically narrower and narrows further with $\bar n$, indicating that the atoms become increasingly localized in a Bragg pattern. The width of this frequency component is determined by the inverse of the mean trapping time, namely, the rate at which jumps between different Bragg gratings occur. 

\begin{figure}[hbtp]
\begin{center}
\includegraphics[width=0.5\textwidth]{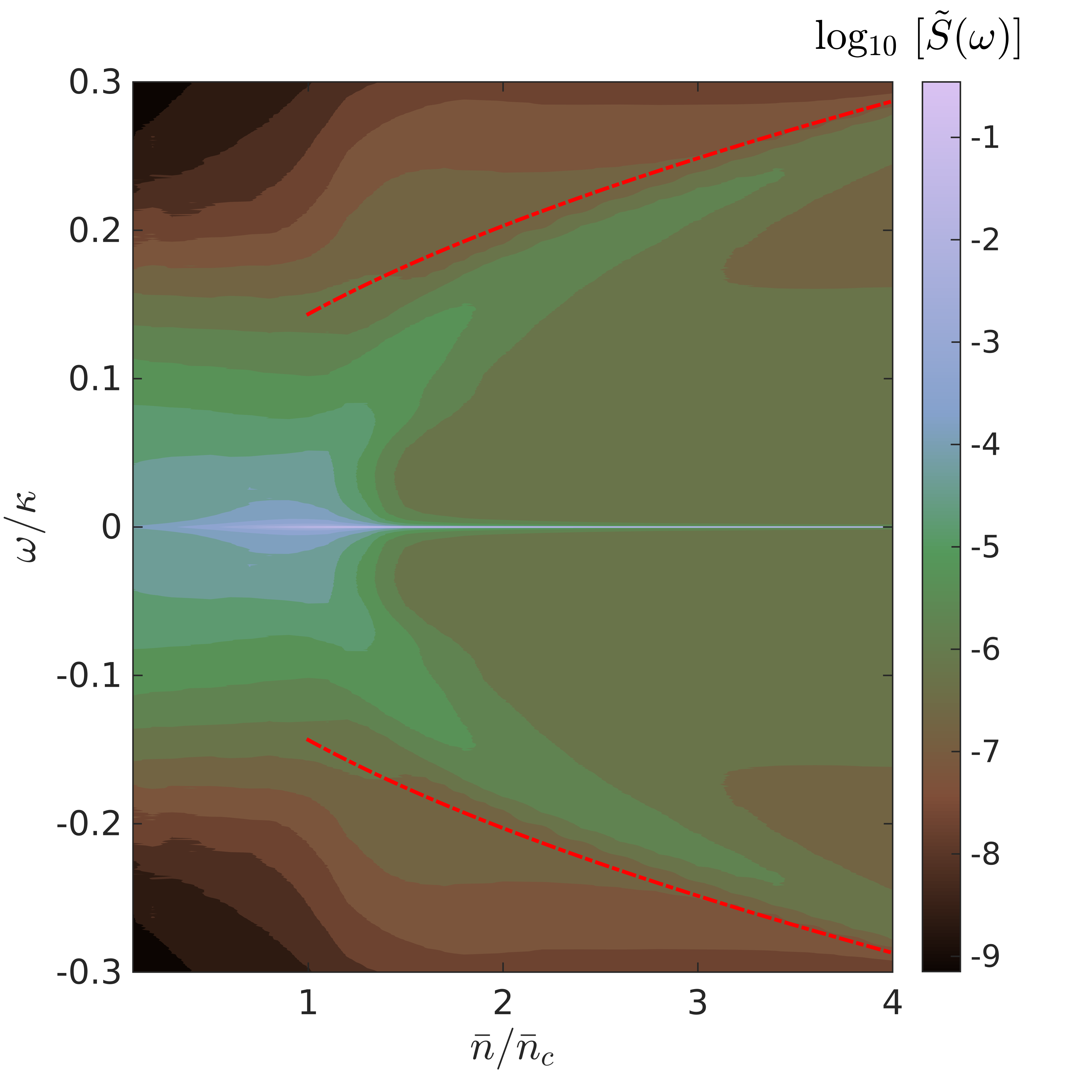}
\caption{(color online) Contour plot of the spectrum of the autocorrelation function $\tilde{S}(\omega)$, Eq. \eqref{S:w}, as a function of $\bar n$ and of the frequency (in units of $\kappa$). The other parameters are the same as in Fig. \ref{fig:cuts}. The red-dashed line corresponds to an estimate deep in the organized regime assuming the atoms are trapped in a harmonic potential with frequency $\tilde{\omega} = \sqrt{2 \omega_r \kappa \bar{n}/\bar{n}_c}$. }
\label{fig:franalysis}
\end{center}
\end{figure}

Above threshold sidebands of the central peak appear, which correspond to the damped oscillations of the autocorrelation function. The central frequency of these sidebands increases for higher pumping strength, while their width decreases.  We understand these features as the onset of oscillations about the minima of the Bragg grating, which one can also observe in the trajectories of Fig. \ref{fig:singleatoms}(a). This conjecture is supported by a simple calculation of the oscillation frequency as a function of $\bar n$, assuming that the potential about their minima is approximated by harmonic oscillators. Even though the estimated frequency is higher, this estimate qualitatively reproduces the dependence of the sidebands central frequency with $\bar n$ above threshold, as visible in Fig. \ref{fig:franalysis}. This plot further shows that the behaviour between the two parameter regions, below and above threshold, are qualitatively very different. The results of our simulations suggest that the transition in Fig. \ref{fig:franalysis} at $\bar n_c$ becomes sharper as the atom number is increased. 

\section{Photon statistics and coherence of the field at the cavity output }
\label{Subsec:cav}

Since the photons scattered by the atoms into the resonator carry the information about the density of the atoms within the cavity spatial mode function, then detection of the light at the cavity output allows to monitor the state of the atoms during the dynamics. This is an established method in experiments with atoms and ions in cavities \cite{Black:2003,Baumann:2011,Brahms:2012,Drewsen,PNAS:2013}, and it is at the basis of proposals for detecting non-destructively the quantum phase of ultracold atoms \cite{Mekhov,Rist}. 

Formally, the field at the cavity output $\hat{a}_{\text{out}} (t)$ is directly proportional to the intracavity field $\hat{a}$ via the relation $\hat{a}_{\text{out}} (t)= \sqrt{2 \kappa} \hat{a} - \hat a_{\text{in}}(t)$, where $\hat a_{\text{in}}(t)$ is the input field, with zero mean value and $[\hat a_{\text{in}}(t),\hat a_{\text{in}}(t')^\dagger]=\delta(t-t')$ \cite{Collett}. The intracavity field is, in turn, given by the solution of the coupled atoms-field dynamics, and under the assumption of time-scales separation it can be cast in the form given in Eq. \eqref{a(t)}, which expresses an effective operator resulting from the coarse-grained dynamics. Equation  \eqref{a(t)} shows that in leading order the intracavity field is proportional to the magnetization $\Theta(t)$, therefore the features of the magnetization we identified this far shall be visible also in the photon statistics at the cavity output. In addition, there is a retardation component, which gives rise to cooling and that in our parameter regime is a small correction. We now report the analysis of the intracavity photon number, and of the first- and second-order correlation functions as a function of the pump strength $\bar n$. Throughout this analysis we will consider that the system has reached the stationary state at $\Delta_c=-\kappa$, corresponding to the minimum temperature of the atoms. Analytically, all averages are taken assuming the atomic distribution is stationary. Numerically, this consists in assuming that the trajectories are evolved starting from the stationary distribution. 

\subsection{Intracavity photon number}

\begin{figure}[hbtp]
\begin{center}
\includegraphics[width=0.48\textwidth]{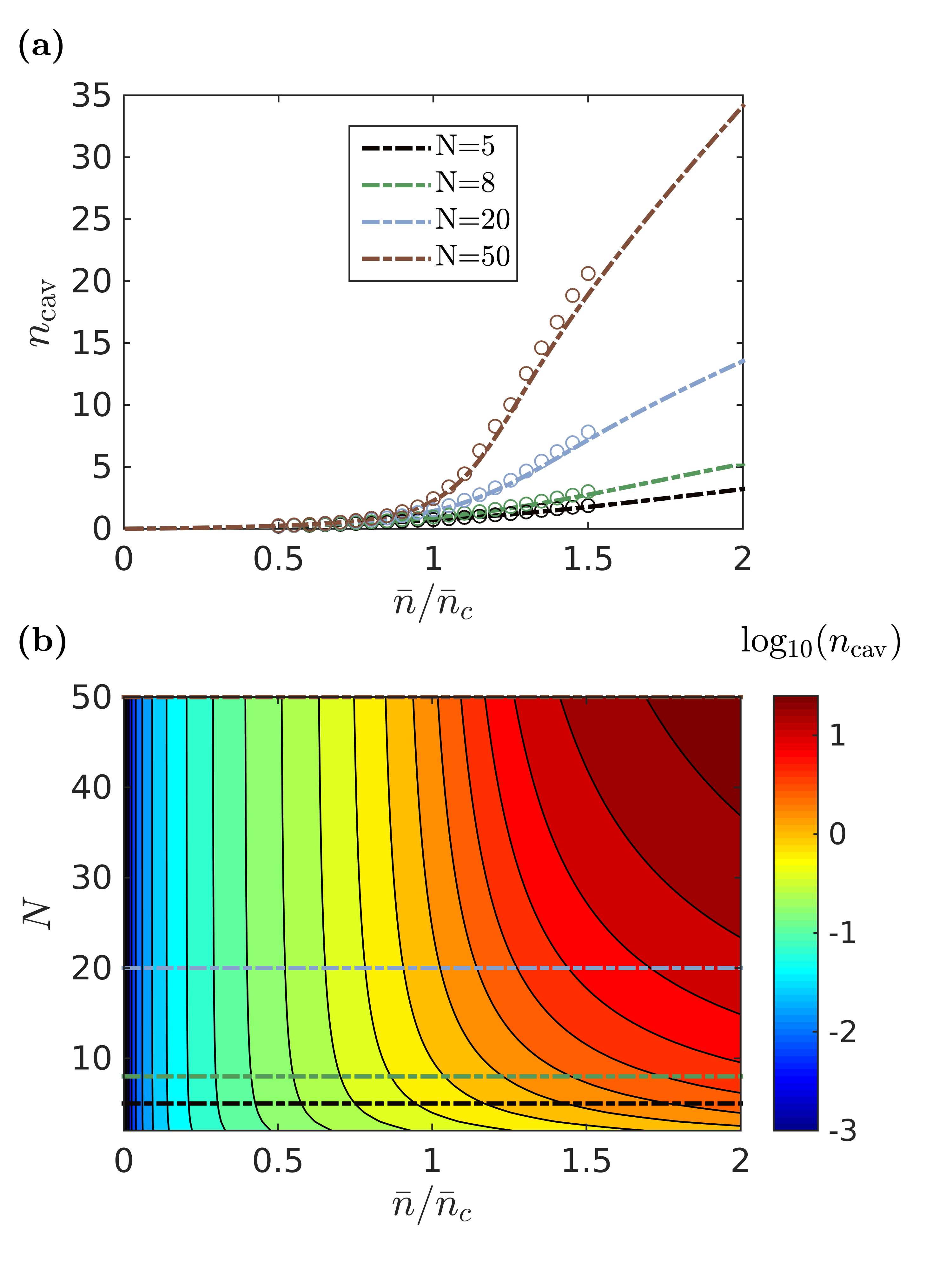}
\caption{
(color online). (a) The mean intracavity photon number $n_{\rm cav}$ at steady state is displayed as a function of the pump strength $\bar n$ (in units of $\bar n_c$) and for different atom numbers (see inset). The circles correspond to the numerical data obtained by using Eq. \eqref{a(t)} and integrating the SDE. The dot-dashed lines correspond to the adiabatic limit $n_{\rm cav}|_{\rm ad}=N\bar n \lim_{t\to\infty} \langle \Theta(t)^2\rangle$, where the average is performed over the stationary state in Eq. \eqref{eq:Steady}. (b) Contour plot of $n_{\rm cav}|_{\rm ad}$ as a function of $N$ and $\bar n$. The colour code is in logarithmic scale. The horizontal lines correspond to the dot-dashed curves shown in subplot (a).} 
\label{fig:g10}
\end{center}
\end{figure}

The intensity of the emitted light is proportional to the mean intracavity photon number
\begin{align} 
n_{\rm cav}=\lim_{t\to\infty}\langle \hat{a}^{\dagger} (t)\hat{a}(t) \rangle \,.
\label{eq:adaggera_0}
\end{align}
Figure \ref{fig:g10} (a) displays $n_{\rm cav}$ as a function of $\bar{n}$ for different atom numbers.
The circles correspond to the mean photon number evaluated by numerical simulations using Eq. \eqref{a(t)}, whereas the dot-dashed lines show the adiabatic solution, Eq. \eqref{photons_adiabatic}, evaluated with the steady-state solution of Eq. \eqref{eq:Steady}. For $\bar n< \bar n_c$ the mean photon number is below unity: Therefore in this regime shot noise is dominant. Above threshold, $n_{\rm cav}$ rapidly increases with $N$ and $\bar n$. For the parameters we choose its value is essentially determined by the adiabatic component of the cavity field, while the contribution due to retardation is negligible (it is less than $0.1 \%$). Thus, the intracavity photon number provides direct access to the autocorrelation function at zero-time delay, $\langle \Theta^2\rangle$.  The numerical data, represented by the circles, follow very closely the curves corresponding to the adiabatic solution $n_{\rm cav}|_{\rm ad}=N\bar n\lim_{t\to\infty} \langle \Theta(t)^2\rangle$. The difference between the two curves is indeed small and due to the effect of the dynamical Stark shift scaling with the parameter $U$, which in the numerics is systematically taken into account. This nonlinear shift of the cavity frequency is maximum when the atoms are localized in a grating and for the chosen sign ($U<0$) it tends to increase the value of $n_{\rm cav}$. 

Figure \ref{fig:g10} (b) displays the contour plot of $n_{\rm cav}$ as a function of $\bar n$ and $N$ using the adiabatic solution, Eq. \eqref{photons_adiabatic}, and the steady-state solution in Eq. \eqref{eq:Steady}.  We observe that well below threshold $n_{\rm cav}$ depends solely on $\bar n$ and is independent of $N$. In this regime, in fact, the atoms are homogeneously distributed, there is no collective effect in photon scattering and thus no superradiance. Using the assumption of a homogeneous spatial distribution and $\bar n\ll \bar n_c$ we can derive an analytical estimate of  $n_{\rm cav}$ which is independent of $N$ (see Appendix  \ref{App:Estimate}): 
\begin{align*}
n_{\rm cav}|_{\bar{n} \ll \bar{n}_c} \approx \bar{n}/2\,.
\end{align*}
As $\bar n$ approaches and then exceeds the threshold value, instead, the dependence of the mean intracavity photon number on $N$ becomes evident. 

\subsection{Spectrum of the emitted light} 

\begin{figure}[hbtp]
\begin{center}
\includegraphics[width=0.48\textwidth]{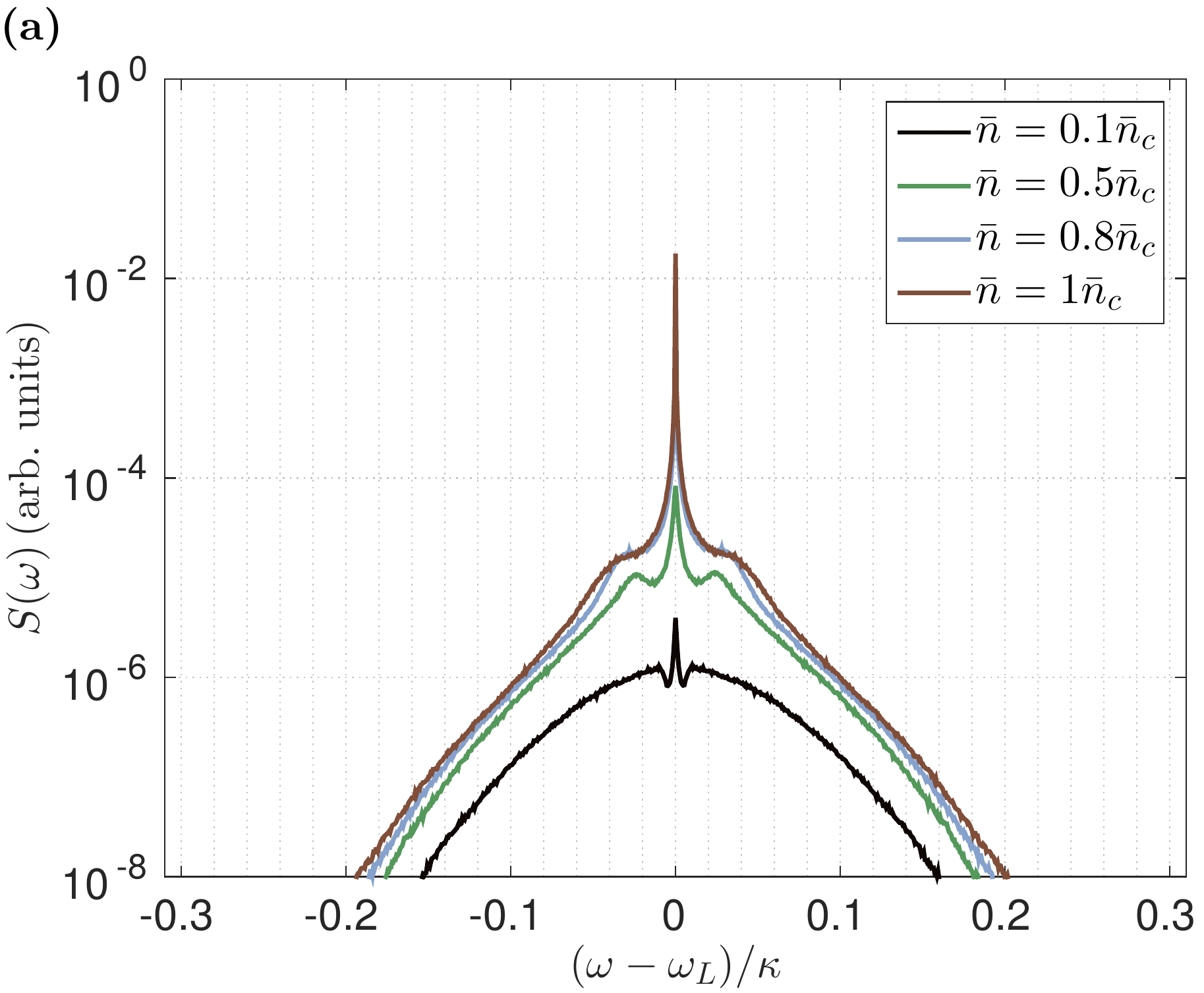}
\includegraphics[width=0.48\textwidth]{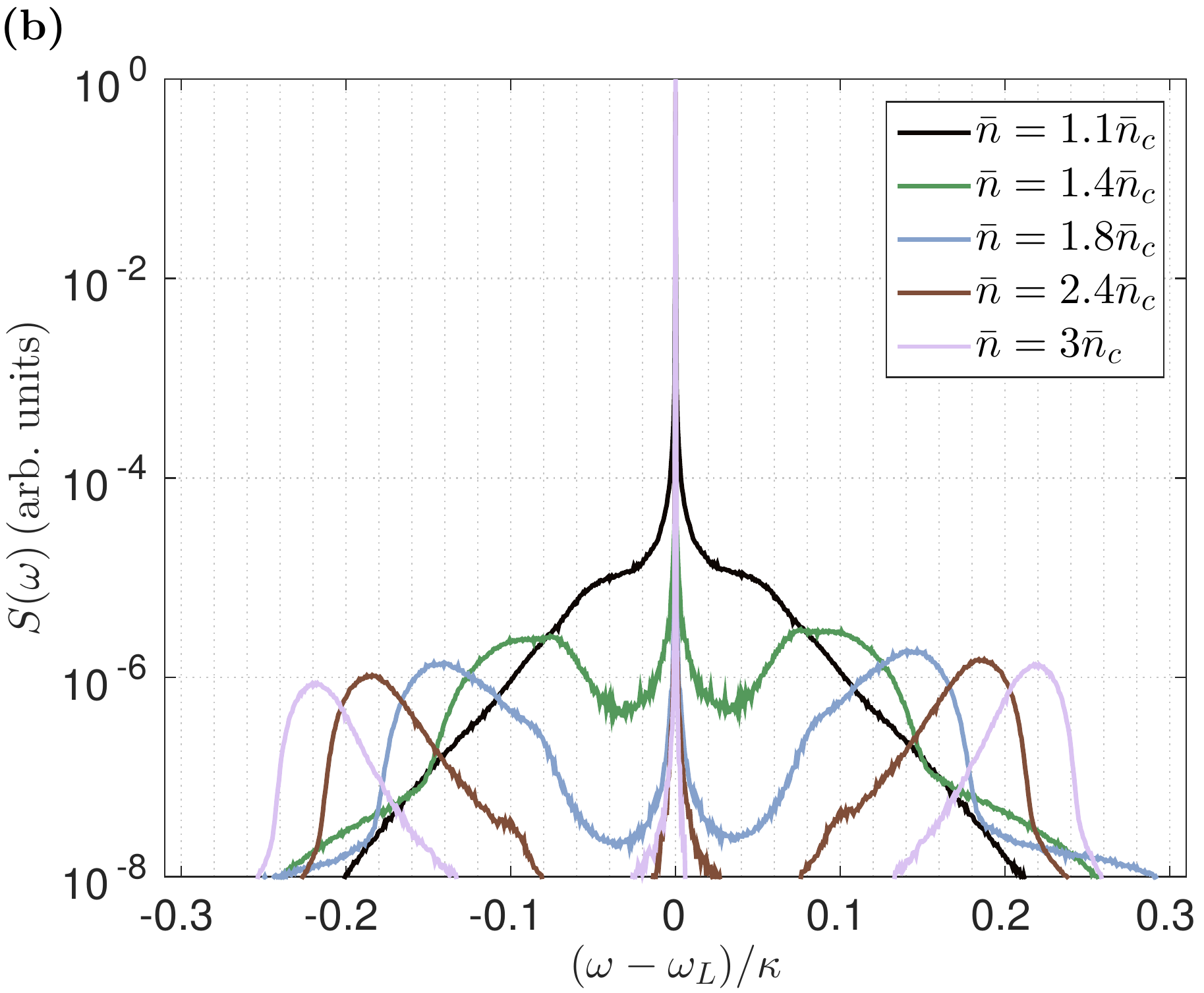}
\caption{(color online) Spectrum of the intracavity field intensity $S(\omega)$, Eq. \eqref{eq:S_omega} and in arbitrary units, at steady state. In (a) the curves correspond to values of $\bar n\le \bar n_c$, and in (b)  to values of $\bar n> \bar n_c$. The data have been numerically evaluated for $N=50$ atoms and over the interval of time $(-10^4:1:10^4) \, \kappa^{-1}$. }
\label{fig:spectrum}
\end{center}
\end{figure}

We now turn to the first-order correlation function at steady state, $g^{(1)}(\tau)=\lim_{t\to\infty}\langle \hat{a}^{\dagger} (t + \tau)\hat{a}(t) \rangle$. At zero-time delay, $\tau=0$, it corresponds to the intracavity photon number. For finite delays $\tau$ it is proportional to the power spectrum of the autocorrelation function. In addition, it contains the nonlinear contribution of the cavity frequency shift and the retarded component of the cavity field. We discuss here the spectrum of $g^{(1)}(\tau)$, 
\begin{align}
S (\omega) &= \lim_{t\to\infty} \frac{1}{2 \pi} \int_{-\infty}^{\infty} {\rm d} \tau e^{- {\rm i} \omega \tau} \langle \hat{a}^{\dagger} (t + \tau) \hat{a} (t) \rangle  \,,
              \label{eq:S_omega}
\end{align} 
which we then compare with the result obtained for the power spectrum of the magnetization. The spectrum $S(\omega)$ is displayed  in Fig. \ref{fig:spectrum} for $N=50$ atoms and different values of the pumping strength.

The behaviour is very similar to the spectrum of the autocorrelation function of the magnetization in Fig. \ref{fig:cuts}. Below threshold, Fig. \ref{fig:spectrum} (a), we observe a broad frequency spectrum, while above threshold, Fig. \ref{fig:spectrum} (b), we notice the emergence of sidebands whose frequency increases with $\bar{n}$. 
In general, the spectrum of the emitted light has the same form as the power spectrum of the magnetization, and thus allows to extract information about the thermodynamics of selforganization.  The contour plot is very similar to the corresponding one of the autocorrelation function, Fig.  \ref{fig:franalysis}. A distinct feature is found in a small asymmetry between the red ($\omega < \omega_L$) and the blue ($\omega > \omega_L$) sideband in Fig. \ref{fig:spectrum} (b).  The asymmetry seems to be due to the contribution of the diabatic component of the cavity field, given in Eq. \eqref{a_retard}. Remarkably, the spectrum qualitatively agrees with  the one observed in experiments analysing selforganization of ultra-cold atoms in single-mode standing-wave resonators \cite{PNAS:2013}, thus outside the regime of validity of the semiclassical treatment. In particular,  sideband asymmetry above threshold was also reported in Ref. \cite{PNAS:2013}. 

\subsection{Intensity-intensity correlations} 

\begin{figure}[hbtp]
\begin{center}
\includegraphics[width=0.54\textwidth]{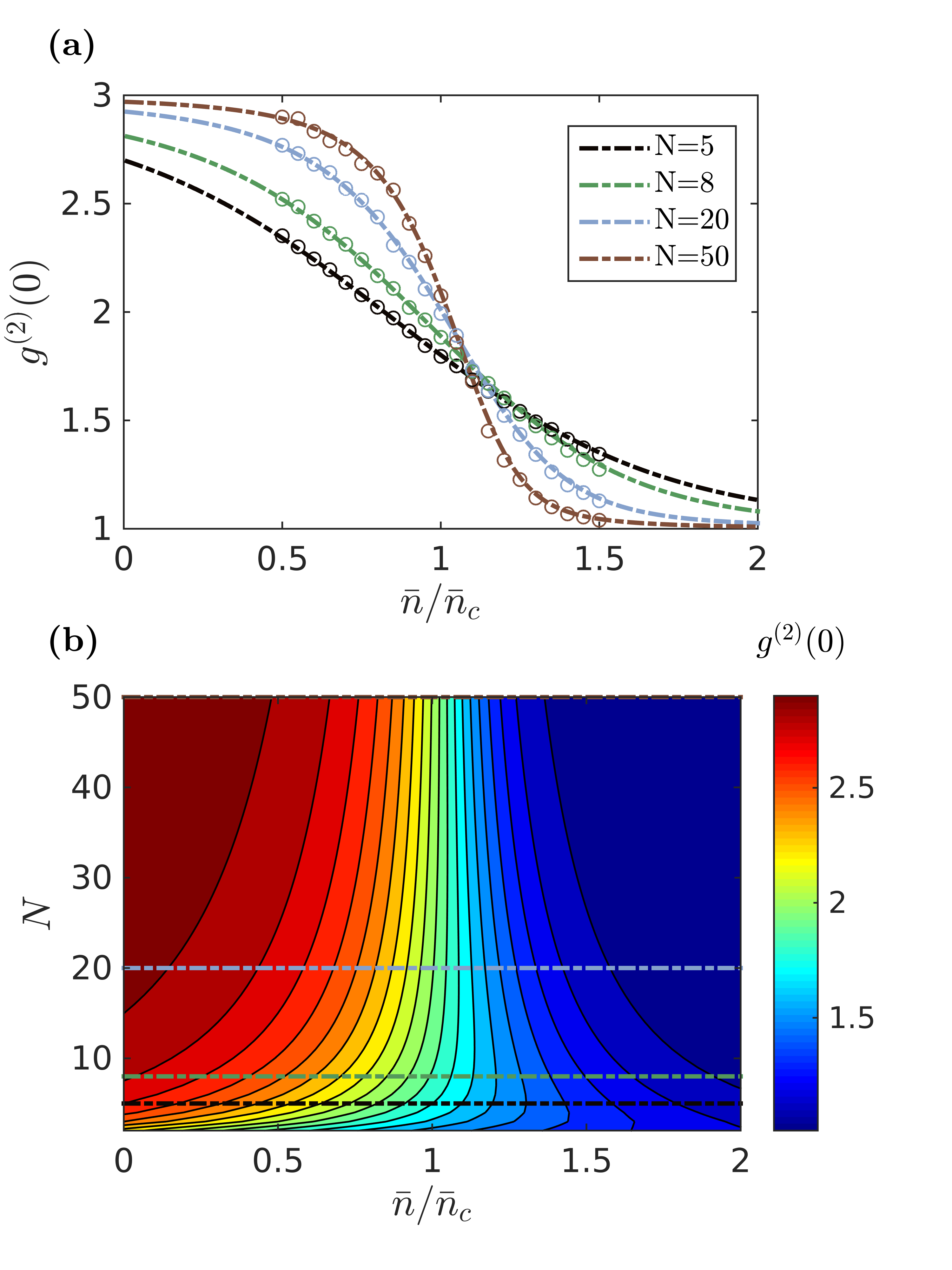}
\caption{
(color online) (a) The intensity-intensity correlation at zero-time delay $g^{(2)} (0)$, Eq. \eqref{eq:int-int-corr}, is shown as a function of the pump strength $\bar{n}$ (in units of $\bar n_c$) and for different atom numbers $N$ (see inset). The circles correspond to the data extracted from numerical simulations, the dot-dashed lines are evaluated using the steady state in Eq. \eqref{eq:Steady} and the adiabatic solution, where the field is proportional to the instantaneous value of the magnetization: $g^{(2)}(0)|_{\rm ad} = \langle \Theta^4 \rangle / \langle \Theta^2 \rangle^2$. (b) Contour plot of the adiabatic component of the intensity-intensity correlation function at zero-time delay $g^{(2)}(0)|_{\rm ad} $ {\it vs.} $\bar n$ and $N$. The horizontal cuts correspond to the dot-dashed lines in subplot (a).}
\label{fig:g20}
\end{center}
\end{figure}

The intracavity photon number below and close to threshold is smaller than unity, and is thus characterized by large photon fluctuations. We now study the properties of these fluctuations by determining the intensity-intensity correlation function,
\beq
g^{(2)}(\tau) =\lim_{t\to\infty} \frac{\langle \hat{a}^{\dagger} (t) \hat{a}^{\dagger} (t+\tau) \hat{a} (t+\tau) \hat{a} (t) \rangle}{\langle \hat{a}^{\dagger} (t) \hat{a} (t) \rangle ^2}\,.
\label{eq:int-int-corr}
\eeq  
with $t \rightarrow \infty$ indicating the steady-state, and focus on its value at zero-time delay, $g^{(2)}(0)$, as a function of $\bar n$  for gaining insight in the photon statistics. Figure \ref{fig:g20} (a) displays the correlation function $g^{(2)}(0)$ as a function of $\bar n$ and for different atom numbers. The circles show $g^{(2)}(0)$ extracted from numerical simulations using Eq. \eqref{a(t)}, while the dot-dashed lines correspond to the adiabatic solution $g^{(2)}(0)|_{\rm ad} = \langle \Theta^4 \rangle / \langle \Theta^2 \rangle^2$ using the steady-state solution in Eq. \eqref{eq:Steady}. Both curves are in good agreement. We observe a crossover from $g^{(2)}(0) \approx 3$ to $g^{(2)} (0) \approx 1$ when tuning the pumping strength from below to above the threshold, which sharpens as $N$ grows. The value above threshold is associated with coherent radiation, which is what one expects when the atoms are locked in a Bragg grating. The behaviour below threshold can be reproduced by means of an analytical model valid for $\bar n\ll \bar n_c$, in the limit in which the atoms form a homogeneous distribution. In Appendix \ref{App:Estimate} we show that in this limit we can write
\begin{align}
g^{(2)}(0) = 3 - 3/(2N),
\label{eq:g20below}
\end{align}
which asymptotically tends to 3 as $N$ increases. This result qualitatively agrees with experimental measurements with ultracold atoms performed below threshold \cite{PNAS:2013}. While this value is also found for squeezed states, in our case we could not find any squeezing in the field quadratures and thus attribute the behaviour of $g^{(2)}(0)$ below threshold to thermal fluctuations.\\
Figure \ref{fig:g20} (b) displays $g^{(2)}(0)$ for different pumping strengths and number of atoms, evaluated using the adiabatic solution $g^{(2)}(0) = \langle \Theta^4 \rangle / \langle \Theta^2 \rangle^2$ and the steady state in Eq. \eqref{eq:Steady}. The dashed horizontal cuts correspond to the dot-dashed curves shown in subplot (a). One clearly observes the crossover from $g^{(2)}(0) \approx 3$ to $g^{(2)}(0) \approx 1$ when $\bar n$ exceeds $\bar{n}_c$, while the transition sharpens for increasing atom numbers.

\section{Conclusions}
\label{Sec:Conclusions}

Atoms can spontaneously form spatially ordered structures in optical resonators when they are transversally driven by lasers. In this paper we have characterized the stationary solution, which emerges from the interplay between the coherent dynamics due to scattering of laser photons into the resonator and the incoherent effects associated with photon losses due to cavity decay. We assumed that these dynamics are characterized by a time-scale separation, such that the cavity field relaxes on a faster time scale to a local steady state depending on the atomic density. This assumption is valid when the cavity loss rate $\kappa$ exceeds the recoil energy $\omega_r$ scaling the mechanical effects of light, and it is fulfilled in several existing experiments \cite{Black:2003,Ritter:2009,Arnold:2012}. Retardation effects are small, but important in order to establish the stationary state. 

Starting from a FPE, which has been derived by means of an ab-initio theoretical treatment \cite{Schuetz:2013}, we have shown that the stationary state is thermal, with a temperature that is solely determined by the detuning between cavity and laser. From this result, we could determine the free energy and thus show that atomic selforganization in a standing-wave cavity mode is  a second-order transition of Landau-type. Our model allows us to determine the phase diagram for the self-organization transition and delivers the critical value of the pump strength in a self-consistent way. This value in agreement with previous estimates \cite{Asboth:2005,Niedenzu:2012}. An interesting further step is to connect this theory with quantum-field theoretical models which analyse selforganization in the ultracold regime \cite{DallaTorre:2011,Piazza:2014,Diehl:2013}, thus extending the validity of our model to the regime in which quantum fluctuations in the atomic motion cannot be treated within a semiclassical model. 

We further remark that, while our analysis focuses on a one-dimensional model, we expect that from our predictions we can extrapolate the stationary behaviour in two spatial dimensions.  This can be calculated by means of a straightforward extension of the treatment in Ref. \cite{Schuetz:2013} to two dimensions. Differing from one dimension, in the symmetry-broken phase the atoms will form a checkerboard pattern as found in Ref. \cite{Baumann:2010}, as long as the atomic gas is uniformly illuminated by the laser and the coupling with the resonator can be treated in the paraxial approximation. The effect of the dimensionality can modify the specific form of friction and diffusion. Moreover, in two dimensions the effect of correlations is expected to be more relevant, so that the statistical properties will be modified.

Photodetection of the emitted light allows one to reveal the thermodynamic  properties of the atoms. Our results show that they exhibit several remarkable analogies with experimental results obtained with ultracold atomic ensembles inside of resonators \cite{PNAS:2013}. While our theory is not generally applicable to these systems, it is not surprising that the field at the cavity output does not depend on the presence (or absence) of matter-wave coherence, as it solely depends on the atomic density. Nevertheless, it would be interesting to identify observables for the cavity field output, if possible, that provide information about quantum coherent properties of matter, in the spirit of matter-wave homodyne detection discussed in Ref. \cite{Rist:2011}. This could be possible when the cavity spectroscopically resolves the many-body excitations, as is verified in the parameter regime of the experimental setup reported in Ref. \cite{Wolke:2012}.

This work is the first of a series analysing the effect of the long-range cavity-mediated interaction. Here we focused on the dynamics at steady state. In Ref. \cite{Jaeger:2015} we will compare the results here reported with a mean-field solution, which is systematically derived from this treatment after making a mean-field ansatz, and discuss its validity in the perspective of developing a BBGKY hierarchy for selforganization in optical resonators \cite{Campa:2009}. In Ref. \cite{Schuetz:2015:2} we will analyse the dynamics of the full distribution after quenches across the phase transition, expanding on the results presented in Ref. \cite{Schuetz:2014}.

\section*{Acknowledgments}
The authors are grateful to R. Landig, G. Manfredi, C. Nardini, F. Piazza, and R. Shaebani for stimulating discussions and helpful comments. This work was supported by the German Research Foundation (DFG, DACH project "Quantum crystals of matter and light") and by the German Ministry of Education and Research (BMBF "Q.com").

\begin{appendix}

\section{Parameters of the Fokker-Planck equation}
\label{App:Parameter}

\noindent In this appendix we give the explicit form of the parameters appearing in the coefficients of Eq. \eqref{L:f}:

\begin{eqnarray}
\delta_F &=& \frac{NU\Theta}{\Delta_c'} \cos(kx_i) \\
\delta_{\Gamma} &= & \cos(kx_j) \frac{NU \Theta}{\Delta_c'} \frac{3 \Delta_c'^2 - \kappa^2}{\Delta_c'^2 
+ \kappa^2}  \\ &&+ \cos(kx_i) \frac{NU \Theta}{\Delta_c'} 
+ 4 \cos(kx_i) \cos(kx_j) \frac{(NU \Theta)^2}{\Delta_c'^2 + \kappa^2} \nonumber\\
\delta_{\eta} &=&  \frac{(2 NU\Theta)^2}{\Delta_c'^2 + \kappa^2}\cos(kx_i) \cos(kx_j)  \\
&&+ \frac{2 NU \Theta \Delta_c'}{-\Delta_c'^2 + \kappa^2} \big\{ \frac{3 \kappa^2 - \Delta_c'^2}{\Delta_c'^2 + \kappa^2}  \cos(kx_j) 
-  \cos(kx_i) \big\}\nonumber\\
\delta_{D} &=& 
 \frac{4 NU\Theta}{\Delta_c'^2 + \kappa^2}  \cos(kx_j)\left(\Delta_c' + \cos(kx_i) NU \Theta\right)
\end{eqnarray}
The diffusion coefficient for the spontaneous decay term reads
\begin{align*}
\mathcal{D}^{\text{sp}}(x_i) &= (\hbar k)^2 \bigg[ \frac{N^2S^2\Theta^2}{\Delta_c'^2+\kappa^2}  [ \sin^2(k x_i) +  \overline{u^2} \cos^2( kx_i) ] \\ 
&+ s  \overline{u^2} (\frac{2NS\Theta \Delta_c'}{\Delta_c'^2+\kappa^2} \cos(k x_i) +s)  \bigg] 
\end{align*}
with $s=\Omega/g$ and $\overline{u^2}$ determines the momentum diffusion due to spontaneous emission recoils projected on the cavity axis (dipole pattern of radiation).

Finally, the correction scaling with $NU/\kappa$ in Eq. \eqref{FPE} reads 
\begin{widetext}
\beq
L_1f=2\hbar k\Delta_c \Theta\sum_i\sin(kx_i)\left[\frac{\Delta_c^2-\kappa^2}{\Delta_c^2+\kappa^2}{\mathcal B}+\Theta\cos(kx_i)\right]\partial_{p_i}f\,
\eeq
\end{widetext}
and is systematically taken into account in our calculations.

%

\section{Stochastic differential Equations}
\label{App:SDE}

The FPE given in Eq. \eqref{FPE} for $|NU| \ll |\Delta_c|$ can be simulated by Stochastic differential equations which in our case read
\begin{align}
\label{SDE:1}
{\rm d}x_{j} &= \frac{p_{j}}{m} {\rm d}t + {\rm d}X_j,\\
{\rm d}p_{j} &= \hbar k  \frac{ 2 S^2 \Delta_c}{\Delta_c^2 + \kappa^2} \sin(k x_j) \bigg( \sum_{i=1}^N \cos(k x_i) \bigg) \delta_U {\rm d}t \label{SDE:2} \\
&+  \frac{8 \omega_r S^2 \Delta_c \kappa}{(\Delta_c^2 + \kappa^2)^2} \sin(kx_j) \bigg( \sum_{i=1}^N \sin(kx_i)  p_i \bigg) {\rm d}t + {\rm d}P_{j}, \nonumber
\end{align}
with
\begin{align}
\delta_U = 1+ \frac{NU}{\Delta_c} \bigg( \frac{\Delta_c^2-\kappa^2}{\Delta_c^2 + \kappa^2} \mathcal{B} + \Theta \cos(kx_j) \bigg),
\end{align}
where $j = 1,...,N$ labels the atoms and ${\rm d}P_j$ denote the momentum noise terms, which are simulated by means of Wiener processes. In particular, $\langle {\rm d}P_j\rangle=0$ and $\langle{\rm d}P_i {\rm d}P_j \rangle= 2 D_{ij} {\rm d}t$ with 
\begin{align}
D_{ij} &= (\hbar k)^2  S^2 \frac{\kappa}{\Delta_c^2 + \kappa^2} \sin(k x_i) \sin(k x_j) 
\end{align}
the element of the diffusion matrix when spontaneous emission is neglected. \\
For $\Delta_c \neq - \kappa$, we additionally take into account position noise ${\rm d}X_i$, which shows cross-correlations with momentum diffusion $\langle{\rm d}P_j {\rm d}X_\ell \rangle =  \eta_{j\ell} {\rm d}t$, with
\begin{eqnarray}
\eta_{j\ell} = 2 \hbar \omega_r S^2 \sin(k x_{j}) \sin(k x_{\ell})  \frac{\kappa^2 - \Delta_c^2}{(\Delta_c^2 + \kappa^2)^2}\,.
\label{eta}
\end{eqnarray}
These terms can only be simulated when adding terms as $\langle {\rm d}X_i {\rm d}X_j \rangle \neq 0$ to the FPE.\\
For the numerical simulations, we use the Heun method \cite{Ruemelin:1982}, which is a second-order Runge Kutta scheme with an Euler predictor.

\section{Determination of the free energy}
\label{App:FreeEnergy}
\noindent The equilibrium state reads
\begin{align}
f({\bm x},{\bm p})=\frac{1}{\mathcal{Z} \Delta^N}\exp\left(-\beta H \right)\,,
\end{align}
with $\mathcal{Z}$ the partition function, $\Delta$ the unit phase space volume, while 
Hamiltonian $H$ is given in Eq. \eqref{Hmf_self}. The canonical partition function $\mathcal{Z}$ takes the form
\begin{align}
\mathcal{Z}&= \left(\frac{\lambda}{\Delta}\right)^N \int_{-1}^{1} {\rm d} \Theta\varOmega(\Theta)
\int_{-\infty}^{\infty} {\rm d}p_1...\int_{-\infty}^{\infty} {\rm d}p_N\exp\left(-\beta H \right) \notag\\
&= \left(\frac{ Z_0\lambda}{\Delta}\right)^N \int_{-1}^{1} {\rm d}\Theta\varOmega(\Theta)\exp\left(-\beta \hbar \Delta_c \bar{n} N \Theta^2\right)\,, \label{Z_app}
\end{align}
with $Z_0 = \sqrt{2m\pi/\beta}$ and
\beq
\varOmega(\Theta)
=\frac{N}{2\pi}\int_{-\infty}^{\infty} {\rm d} \omega \exp\left(i\omega N\Theta\right)J_0(\omega)^N\,,
\label{omega_theta}
\eeq
where $J_n(w)=
1/(i^n \lambda)
\int_{0}^{\lambda} {\rm d}x\cos(nkx)\exp(i\omega\cos(kx))$ is the $n$-th order Bessel function \cite{Abramowitz-Stegun}. 
In order to compute Eq. \eqref{omega_theta}, we rewrite it as
\begin{align}
\varOmega(\Theta)
&=\frac{N}{2\pi}\int_{-\infty}^{\infty} d\omega \exp\left( N h(\omega)\right)\,, \label{omega_h} 
\end{align}
where we introduced the function
\begin{align}
h(\omega) &= i\omega \Theta+\log\left(J_0(\omega)\right). 
\label{h_omega}
\end{align}
We can now compute the integral in Eq. \eqref{omega_h} using the method of steepest descent. For this purpose, we derive the stationary condition for Eq. \eqref{h_omega}. This reads 
\begin{align*}
i\Theta-\frac{ J_1(\omega_0) }{J_0(\omega_0)}=0\,,
\end{align*}
which we can rewrite as 
\begin{align}
\Theta= q(\gamma_0)=\frac{I_1(\gamma_0) }{I_0(\gamma_0)}
\end{align}
after defining $\omega_0=i\gamma_0$ and using that  $\frac{ J_1(\omega_0) }{J_0(\omega_0)}=i\frac{I_1(\gamma_0) }{I_0(\gamma_0)}$.
The function
$q:\mathbb{R}\to(-1,1)$
with
 $  y\mapsto    \frac{ I_1(y) }{I_0(y)}$    
is 
bijective, 
such that there is a unique solution satisfying the equation
\begin{align}
\gamma_0=q^{-1}(\Theta).
\label{Eq:gamma_0}
\end{align}
With the method of steepest descent, we get
\begin{align}
 &\varOmega(\Theta) 
 \notag \sim\frac{N}{2\pi}\sqrt{\frac{2\pi}{N |h''(\omega_0)|}}\exp \Big[ Nh(\omega_0) \Big] \\
&=\sqrt{\frac{N}{2\pi}} C(\Theta)
\exp \Big[ N \big\{\log\left(I_0(q^{-1}(\Theta))\right)-q^{-1}(\Theta)\Theta \big\} \Big] \label{omega_largeN}
\end{align}
with $$C(\Theta) = \left|\Theta^2-\frac{I_0(q^{-1}(\Theta))+I_2(q^{-1}(\Theta))}{2I_0(q^{-1}(\Theta))}\right|^{-\frac{1}{2}}\,.$$
Using Eq. \eqref{omega_largeN} in Eq. \eqref{Z_app}, at leading order in $N$ we can cast the canonical partition function into the form
\begin{align}
 \mathcal{Z}&= \left(\frac{Z_0\lambda}{\Delta}\right)^N  \int_{-1}^{1}d\Theta\sqrt{\frac{N}{2\pi}} 
 C(\Theta)
 \exp(-\beta N\mathcal{F}(\Theta)) \notag\,,
\end{align}
where $\mathcal{F}({\Theta})$ is the free energy per particle, 
\begin{align}
\beta ( \mathcal{F}(\Theta) - \mathcal{F}_0 )= \beta \hbar \Delta_c \bar{n} \Theta^2 +q^{-1}(\Theta)\Theta-\log\left(I_0(q^{-1}(\Theta))\right)\,,
\label{Eq:betaF}
\end{align}
and $- \beta N \mathcal{F}_0 =N \log (Z_0 \lambda / \Delta )$. After performing a Taylor expansion of Eq. \eqref{Eq:betaF} for small values of the order parameter, close to $\Theta=0$, we obtain
\begin{align}
\beta \big( \mathcal{F} (\Theta) - \mathcal{F}_0 \big) \approx (1- \bar{n}/\bar{n}_c) \Theta^2 + \frac{1}{4} \Theta^4,
\end{align}
which shows that close to the instability the free energy can be cast into the form of a Landau potential \cite{Landau}. This shows that the system undergoes, in the considered limit, a second order phase transition at the critical value $\bar{n}=\bar{n}_c$
with
\beq
\bar{n}_c = \frac{\kappa^2 + \Delta_c^2}{4 \Delta_c^2}.
\eeq
We use the method of steepest descent to minimize $ \mathcal{F}(\Theta)$ in Eq. \eqref{Eq:betaF}
and find that the free energy is stationary if the order parameter solves the equation:
\beq
\Theta = q \left(2 \frac{\bar{n}}{\bar{n}_c} \Theta \right).
\label{eq:thetastat}
\eeq 

\section{Analytical estimates}
\label{App:Estimate}

Several quantities of relevance can be analytically determined in the limit of small pumping strength, specifically when $\bar{n} \ll \bar n_c$. In this limit we assume that the atoms move ballistically and their spatial distribution is homogeneous. The steady state then reads
\begin{align*}
f_s(\bm{x},\bm{p}) = \frac{1}{\lambda^N} \Big( \frac{\beta}{2 \pi m} \Big)^{N/2} \exp\Big( - \beta \sum_i \frac{p_i^2}{2m} \Big),
\end{align*}
which is a homogeneous distribution for the atoms, while the momentum distribution is thermal with $\beta$ defined in Eq. \eqref{eq:hbar_beta}. The mean value of the order parameter for this distribution vanishes $\langle \Theta \rangle = 0$,
while fluctuations scale as
\begin{align}
\langle \Theta^2 \rangle = \int {\rm d} \bm{x} \int {\rm d} \bm{p} f_s(\bm{x},\bm{p}) 
\Theta^2 = \frac{1}{2N}.
\label{est_theta2}
\end{align}
Here we used that the cross-terms in $\Theta^2 = \sum_{i,j} \cos(kx_i) \cos(kx_j) / (N^2)$ vanish for a homogeneous distribution. For the standard deviation $\Delta \Theta = \big( \langle \Theta^2 \rangle - \langle \Theta \rangle^2\big)^{1/2}$ we thus find
\begin{align}
\Delta \Theta = \sqrt{\frac{1}{2N}}
\label{Delta Theta}
\end{align}
which shows that the width $\Delta \Theta_0$ for the distribution function $P_N(\Theta_0)$ in Eq. \eqref{eq:PTheta_0} decreases with $N^{-1/2}$ for very low pumping strengths. 
We checked that for $\bar{n} \ll \bar{n}_c$ the Gaussian assumption is a good approximation for low values of $|\Theta_0|$  and sufficiently large atom number. This result is reported in Eq. \eqref{P:theo}.\\
In section \ref{Subsec:cav} cavity field properties such as mean photon number $\langle \hat{a}^{\dagger} \hat{a} \rangle$ and intensity-intensity correlations at zeros time delay $g^{(2)} (0)$ are discussed. By adiabatically eliminating the cavity field, i.e. using Eq. \eqref{photons_adiabatic}, and neglecting the dynamical Stark shift, we can give the following estimate for the mean photon number
\begin{align}
\langle \hat{a}^{\dagger} \hat{a} \rangle = N \bar{n} \langle \Theta^2 \rangle = \bar{n} / 2 = \frac{\bar{n}_c}{2} \frac{\bar{n}}{\bar{n_c}}
\label{est_meanphoton}
\end{align}
under the assumption of a homogeneous spatial distribution. As long as the spatial distribution remains homogeneous, the mean photon number thus scales with the ratio $\bar{n} / \bar{n}_c$ independent on the atom number $N$. This result is discussed in Sec. \ref{Subsec:cav} A and gets evident in Fig. \ref{fig:g10} (b). 
Under the same conditions, far below threshold, we get
\begin{align}
\langle \Theta^4 \rangle &=\int {\rm d} \bm{x} \int {\rm d} \bm{p}\, f_s(\bm{x},\bm{p}) \left( \sum_i \cos(kx_i) / N \right)^4
\label{est_theta4}
\\
&= \frac{1}{N^4} \left( N \frac{I_{(4)}}{2\pi} + 3 N(N-1) \frac{ I_{(2)}^2 }{(2 \pi)^2} \right) 
= \frac{3(N-1)}{8N^3}
\nonumber
\end{align} 
with $I_{(2)} = \int_0^{2 \pi} {\rm d} \tilde{x} \cos^2(\tilde{x})$ and $I_{(4)} = \int_0^{2 \pi} {\rm d} \tilde{x} \cos^4(\tilde{x})$. 
For the intensity-intensity correlations at zero time delay
\begin{align}
g^{(2)}(0) = \langle \Theta^4 \rangle / \langle \Theta^2 \rangle ^2,
\label{est_g20}
\end{align}
using Eqs. \eqref{est_theta2} and \eqref{est_theta4}, we thus find
\begin{align}
\lim_{ \bar{n} \rightarrow 0 } g^{(2)}(0) = 3 - \frac{3}{2N}.
\label{g20_estimate}
\end{align}
This function tends towards the value of 3 for increasing atom numbers, as can be seen in Fig. \ref{fig:g20}.\\
When assuming ballistic expansion, which is justified whenever the forces on the atoms due to cavity backaction are small, i.e. far below threshold, we can also derive an analytical estimate for the correlation function $C(\tau) = \langle \Theta (t) \Theta(t+\tau)\rangle$ at steady state
\begin{align}
\label{eq:corrbelow}
&\lim_{\bar{n} \rightarrow 0} \langle \Theta (t) \Theta(t+\tau) \rangle \\
= &\langle \Theta^2 \rangle_t \Big( \frac{\beta}{2 \pi m}\Big)^{1/2} \int {\rm d} p \exp\Big( -\beta \frac{p^2}{2m}\Big) \cos\Big(k \frac{p}{m} \tau\Big) \nonumber \\
= & \langle \Theta^2 \rangle_t \exp \Big( - \frac{\omega_r}{\hbar \beta} \tau^2\Big) = \langle \Theta^2 \rangle_t \exp\Big( - (\tau/ \tau_c^{\rm free})^2\Big) \nonumber
\end{align}
with $\tau_c^{\rm free} = \sqrt{(\hbar \beta / \omega_r)}$, where $\beta$ is the inverse temperature defined in Eq. \eqref{eq:hbar_beta} and $\langle \Theta^2 \rangle_t = \frac{1}{2N}$ according to Eq. \eqref{est_theta2}. The result is reported in Eq. \eqref{eq:corr:theo}.

\end{appendix} 

\end{document}